\documentclass[useAMS,letterpaper,usenatbib, onecolumn]{mn2e}
\pdfoutput=1
\usepackage{graphicx}
\usepackage{array,amssymb,amsmath}
\usepackage[usenames]{color}
\usepackage{mathrsfs}
\usepackage{tikz}
\newcommand{\nc}{\newcommand}

\newcommand{\be}{\begin{equation}}
\newcommand{\ee}{\end{equation}}
\newcommand{\bea}{\begin{eqnarray}}
\newcommand{\eea}{\end{eqnarray}}

\renewcommand\[{\left[}
\renewcommand\]{\right]}

\newcommand\Mpc{\,\mbox{Mpc}}

\newcommand\lsim{\mathrel{\rlap{\lower4pt\hbox{\hskip1pt$\sim$}}
    \raise1pt\hbox{$<$}}}
\newcommand\gsim{\mathrel{\rlap{\lower4pt\hbox{\hskip1pt$\sim$}}
    \raise1pt\hbox{$>$}}}
\def\ee{\end{equation}}
\def\be{\begin{equation}}
\newcommand{\Omk}{\Omega_\kappa}

\newcommand{\Om}{\Omega_m}
\newcommand{\OL}{\Omega_\Lambda}

\newcommand{\mc}[1]{{\mathcal{#1}}}

\newcommand{\dr}{\text{d}}

\newcommand{\Ode}{\Omega_\text{de}}

\newcommand{\Cp}{{\mathscr{C}}}

\newcommand{\data}{D}
\newcommand{\mb}{m_B^*}
\newcommand{\mbi}{m_{Bi}^*}
\newcommand{\mbih}{\hat{m}_{Bi}^*}
\newcommand{\norm}{\mathcal{N}}
\newcommand{\like}{{\mathscr{L}}}

\newcommand{\diag}[1]{\text{diag} \! \left( {#1} \right)}

\newcommand{\sx}{\sigma_x}
\newcommand{\sy}{\sigma_y}

\newcommand{\sxs}{\sigma_x^2}
\newcommand{\sys}{\sigma_y^2}

\newcommand{\xh}{\hat{x}}
\newcommand{\yh}{\hat{y}}

\newcommand{\mhi}{\hat{m}_{Bi}^*}

\newcommand{\Nd}{\mathcal{N}}

\newcommand{\mbl}{\underline{m}^*_B}

\newcommand{\bub}{\underline{b}}
\newcommand{\Zm}{\Sigma_m}

\newcommand{\Zz}{\Sigma_z}

\newcommand{\Rs}{R_x}
\newcommand{\Rc}{R_c}

\newcommand{\Real}{\mathbb{R}}
\newcommand{\Zd}{\Sigma_{\Delta}}
\newcommand{\mg}{{\underline{M}}}
\newcommand{\muu}{\underline{\mu}}

\newcommand{\xbh}{\hat{\underline{x}}_1}

\newcommand{\cbh}{\hat{\underline{c}}}
\newcommand{\zbh}{\hat{\underline{z}}}
\newcommand{\mbh}{\hat{\underline{m}}_B^*}
\newcommand{\cbl}{\underline{c}}
\newcommand{\zbl}{\underline{z}}

\newcommand{\xbl}{\underline{x}_1}

\newcommand{\chisq}{\chi^2}
\newcommand{\salt}[1]{{\sc SALT-II}}
\newcommand{\mlcs}[1]{{\sc MLCS}}

\newcommand{\Scov}{\Sigma_C}
\newcommand{\hf}{\frac{1}{2}}
\newcommand{\Sa}{\Sigma_A}
\newcommand{\Sigp}{\Sigma_P}
\newcommand{\So}{\Sigma_0}

\newcommand{\jj}{{\underline{J}}}

\nc{\Cm}{\hat{C}}
\nc{\sigint}{\sigma_{\mu}^\text{int}}
\nc{\muth}{\mu^{\text{th}}}
\nc{\cnot}{c_\star}
\nc{\xnot}{x_\star}
\nc{\onesn}{\mathbf{1}_n}
\nc{\ul}[1]{\underline{#1}}
\nc{\diff}{{\mathcal{T}}}
\nc{\lcdm}[1]{$\Lambda$CDM}

\begin{document}
\setlength{\unitlength}{1mm}
%\twocolumn[\hsize\textwidth\columnwidth\hsize\csname@twocolumnfalse\endcsname]
%\title[Improved cosmological parameters constraints]{Improved cosmological parameters constraints from SNIa data}
%CURRENT ASTRO TITLE
\title[Improved constraints from SNIa]{Improved constraints on cosmological parameters from SNIa data}
%STATS TITLE 
%\title[Bayesian hierarchical model for SNIa]{A Bayesian hierarchical model for cosmological parameter inference from supernovae type Ia data}

%Stronger cosmological constraints from SNIa data

\author[March et al]{M.C. March$^{1}\thanks{Corresponding author:
marisa.march06@imperial.ac.uk}$, R.~Trotta$^{1,2}$, P.~Berkes$^{3}$, G.D.~Starkman$^{4}$,  and P.M.~Vaudrevange$^{4,5}$\\
$^{1}$Imperial College London, Astrophysics Group, 
	  Blackett Laboratory, Prince Consort Road, London SW7 2AZ, UK\\
$^{2}$African Institute for Mathematical Sciences, 6 Melrose Rd, Muizenberg, 7945, Cape Town, South Africa\\
$^{3}$Brandeis University, Volen Center for Complex Systems, 415 South Street, Waltham, MA 02454-9110, USA \\
$^{4}$CERCA \& Department of Physics, Case Western Reserve University, 10900 Euclid Ave, Cleveland, OH 44106, USA \\
$^{5}$DESY, Notkestrasse 85, 22607 Hamburg, Germany
}
\date{\today, preprint: DESY-11-023}

\maketitle
%\begin{keywords}
%Preprint number: 
%\end{keywords}

\begin{abstract}
We present a new method based on a Bayesian hierarchical model to extract constraints on cosmological parameters from SNIa data obtained with the \salt{} lightcurve fitter. We demonstrate with simulated data sets that our method delivers tighter statistical constraints on the cosmological parameters over 90\% of the time, that it reduces statitical bias typically by a factor $\sim 2-3$ and that it has better coverage properties than the usual $\chisq$ approach. As a further benefit, a full posterior probability distribution for the dispersion of the intrinsic magnitude of SNe is obtained. We apply this method to recent SNIa data, and by combining them with CMB and BAO data we obtain $\Om=0.28 \pm 0.02, \OL=0.73 \pm 0.01$ (assuming $w=-1$) and $\Om=0.28 \pm 0.01, w=-0.90 \pm 0.05$ (assuming flatness; statistical uncertainties only).  
We constrain the intrinsic dispersion of the B-band magnitude of the SNIa population, obtaining $\sigint = 0.13 \pm 0.01$ [mag]. Applications to systematic uncertainties will be discussed in a forthcoming paper.
\end{abstract}
\bigskip

\maketitle

\section{Introduction}

Since the late 1990s when the Supernova Cosmology Project and the High-Z Supernova Search Team presented
evidence that the expansion of the universe is accelerating~\citep{Riess:1998cb,Perlmutter:1998np}, 
observations of Type Ia supernovae (SNIa) has been seen as one of our most important tools for measuring the
cosmic expansion as a function of time.  Since a precise measurement of the evolution of the scale factor is 
likely the key to  characterizing the dark energy or establishing that General Relativity must be modified on 
cosmological scales, the limited data that our universe affords us must be used to the greatest 
possible advantage.   One important element of that task is the most careful possible statistical analysis of the data.
Here we report how an improved statistical approach to SNIa data, 
with, in particular, more consistent treatment of uncertainties, 
leads to significant improvements in both the precision and accuracy of inferred cosmological parameters.

The fundamental assumption underlying the past and proposed use of Type Ia supernovae to measure the 
expansion history is that they are ``standardizable candles".   Type Ia supernovae occur when material
accreting onto a white dwarf from a companion drives the mass of the white dwarf above the maximum
that can be supported by electron degeneracy pressure, the Chandrasekhar limit of about $1.4$ solar masses.
This triggers the collapse of the star and  the explosive onset of carbon fusion, 
in turn powering the supernova explosion.   
Because the collapse happens at a particular critical mass, all Type Ia supernovae are similar.
Nevertheless, variability in several factors, including composition, rotation rate, and accretion rate,
can lead to measurable differences in supernova observables as a function of time.  
Indeed,  the intrinsic magnitude of the nearby  Type Ia supernovae,
the distances to which are known via independent means,
exhibit a fairly large scatter. 
Fortunately, this scatter can be reduced by applying the so-called ``Phillips corrections'' --
phenomenological correlations between the intrinsic magnitude of SNIa and their colour 
as well as between their intrinsic magnitudes and the time scale for the decline of their luminosity~\citep{Phillips:1993ng,Phillips:1999vh}. 
Such corrections are derived from multi-wavelength observations of the SNIa lightcurves 
(i.e., their apparent brightness as a function of time).  
Fortuitously, they make SNIa into standardizable candles -- in other words, having measured the 
colour and light curve of a SNIa, one can infer its intrinsic magnitude with relatively low scatter, typically in the range $0.1-0.2$ mag.
Observations of SNIa at a range of redshifts can then be used to measure the evolution of luminosity distance
as a function of redshift, and thence infer the evolution of the scale factor,   assuming that the intrinsic
properties of SNIa do not themselves evolve (an assumption that has to be carefully checked). 

The SNIa sample which is used to measure distances in the Universe has grown massively thanks to a world-wide observational effort~\citep{Astier:2005qq,WoodVassey2007,Amanullah2010Spectra,Kowalski2008Improved,Kessler2009Firstyear,Freedman:2009vv,Contreras2009Carnegie,Balland2009ESOVLT,Bailey2008Initial,Hicken2009CfA3}. Presently, several hundred  SNIa have been observed, 
a sample which is set to increase by an order of magnitude in the next 5 years or so. 
As observations have become more accurate and refined, discrepancies in their modeling have come into focus.
Two main methods have emerged to perform the lightcurve fit and derive cosmological parameter constraints.
The Multi-Colour Lightcurve Shape (\mlcs{}) \citep{JhaRiess2007} strategy is to simultaneously infer the Phillips corrections and the cosmological parameters of interest, 
applying a Bayesian prior to the parameter controlling extinction.
The SALT and \salt{} \citep{GuyAstier2007} methodology splits the process in two steps. 
First, Phillips corrections are derived from the lightcurve data;
the cosmological parameters are then constrained in a separate inference step. 
As the supernova sample has grown and  improved, 
the tension between the results of the two methods has increased.

Despite the past and anticipated improvements in the supernova sample, 
the crucial inference step of deriving cosmological constraints from the \salt{} lightcurve
fits has remained largely unchanged. For details of how the cosmology fitting is currently done, see for example \citep{Astier:2005qq,Kowalski2008Improved,Amanullah2010Spectra,Conley:2011ku}.  As currently used, it 
suffers from several shortcomings, such as not allowing for rigorous model checking, 
and not providing a rigorous framework for the evaluation of systematic uncertainties. 
The purpose of this paper is to introduce a statistically principled, rigorous, Bayesian hierarchical model for cosmological parameter fitting to SNIa data from \salt{} lightcurve fits. 
In particular the method addresses  identified shortcomings of standard chi-squared approaches  --
notably, it  properly accounts for the dependence of the errors in the distance moduli of the supernovae on 
fitted parameters.  It also treats more carefully  the uncertainty on the Phillips colour correction parameter,
escaping the pontetial bias caused by the fact that the error
is comparable to the width of the distribution of its value.

We will show that our new method delivers considerably tighter statistical constraints on the parameters of interest, 
while giving a framework for the full propagation of systematic uncertainties to the final inferences. 
(This will be explored in an upcoming work.) 
We also apply our Bayesian hierarchical model to current SNIa data, 
and derive new cosmological constraints from them. We derive the intrinsic scatter in the SNIa absolute magnitude and obtain a statistically sound uncertainty on its value.

This paper is organized as follows: in section \ref{sec:bhm_top} we review the standard method used to perform cosmological fits from \salt{} lightcurve results and we describe its limitations. We then present a new, fully Bayesian method, which we test on simulated data in section~\ref{sec:tests}, where detailed comparisons of the performance of our new method with the standard approach are presented. We apply our new method to current SN data in section~\ref{sec:current} and give our conclusions in section~\ref{sec:conclusions}. 

\section{Cosmology from \salt{} lightcurve fits}
\label{sec:bhm_top}

\subsection{Definition of the inference problem}

Several methods are available to fit SNe lightcurves, including the \mlcs{} method, the $\Delta m_{15}$ method, CMAGIC, \citep{WangGoldhaber2003,ConleyGoldhaber2006} SALT, \salt{} and others. Recently, a sophisticated Bayesian hierarchical method to fit optical and infrared lightcurve data has been proposed by~\cite{Mandel:2009xr,Mandel:2010xj}. As mentioned above,  MLCS fits the cosmological parameters at the same time as the parameters controlling the lightcurve fits. The SALT and \salt{} methods, on the contrary, first fit to the SNe lightcurves three parameters controlling the SN magnitude, the stretch and colour corrections. From those fits, the cosmological parameters are then fitted in a second, separate step. In this paper, we will consider the \salt{} method (although our discussion is equally applicable to SALT), and focus on the second step in the procedure, namely the extraction of cosmological parameters from the fitted lightcurves. We briefly summarize below the lightcurve fitting step, on which our method builds.
 
The rest-frame flux at wavelength $\lambda$ and time $t$ is fitted with the expression
\be
\frac{d F_{\rm rest}}{d \lambda} (t, \lambda) = x_0 \left[M_0(t, \lambda) + x_1 M_1(t, \lambda)  \right] \exp\left(c \cdot CL(\lambda) \right),
 \ee
where $M_0, M_1, CL$ are functions determined from a training process, while the fitted parameters are $x_0$ (which controls the overall flux normalization), $x_1$ (the stretch parameter) and $c$ (the colour correction parameter). The $B$-band apparent magnitude $\mb$ is related to $x_0$ by the expression
\be
\mb = -2.5 \log\left[ x_0 \int d\lambda M_0(t=0, \lambda) T^B(\lambda) \right],
\ee
where $T^B(\lambda)$ is the transmission curve for the observer's $B$-band, and $t=0$ is by convention the time of peak luminosity. After fitting the SNIa lightcurve data with \salt{} algorithm, e.g.~\cite{Kessler2009Firstyear} report the best-fit values for $\mb, x_1, c$, the best-fit redshift $z$ of each SNIa and a covariance matrix $\Cm_i$ for each SN, describing the covariance between $\mb, x_1, c$ from the fit, of the form
\be \label{eq:def_Cm}
\Cm_i = \left( \begin{array}{c c c} \sigma_{\mb i}^2  & \sigma_{\mb i,  x_1i} & \sigma_{\mb i,  ci}  \\ 
\sigma_{\mb i,  x_1i} & \sigma^2_{x_1i} & \sigma_{x_1i,ci}  \\
\sigma_{\mb i,  ci} & \sigma_{x_1i,ci} & \sigma^2_{ci}  
\end{array} \right). 
\ee
Let us denote the result of the \salt{} lightcurve fitting procedure for each SN as 
\be \label{eq:data}
\data_i = \{ \hat{z}_i, \hat{m}_{Bi}^*, \hat{x}_{1i},\hat{c}_i, \Cm_i \} .
\ee
(where $i$ runs through the $n$ SNe in the sample, and measured quantities are denoted by a hat). We assume (as it is implicitly done in the literature) that the distribution of $ \hat{m}_{Bi}^*, \hat{x}_{1i},\hat{c}_i$ is a multi-normal Gaussian with covariance matrix $\Cm_i$. 

The distance modulus $\mu_i$ for each SN (i.e., the difference between its apparent B--band magnitude and its absolute magnitude) is modeled as:
\be \label{eq:observed_distance_modulus}
\mu_i = \mbi - M_i + \alpha \cdot x_{1i} - \beta \cdot c_i
\ee
where $M_i$ is the (unknown) $B$-band absolute magnitude of the SN, while $\alpha, \beta$ are nuisance parameters controlling the stretch and colour correction (so-called ``Phillips corrections''), respectively, which have to be determined from the data at the same time as the parameters of interest. The purpose of applying the Phillips corrections is to reduce the scatter in the distance modulus of the supernovae, so they can be used as almost standard candles. However, even after applying the corrections, some intrinsic dispersion in magnitude is expected to remain. Such intrinsic dispersion can have physical origin (e.g., host galaxies properties such as mass~\citep{Kelly:2009iy,Sullivan:2011kv} and star formation rate~\citep{Sullivan:2006ah}, host galaxy reddening~\citep{Mandel:2010xj}, possible SNe Ia evolution~\citep{GonzalezGaitan:2010iw}, etc) or be associated with undetected or underestimated systematic errors in the surveys. Below, we show how to include the intrinsic dispersion explicitly in the statistical model.
 
Turning now to the theoretical predictions, the cosmological parameters we are interested in constraining are 
\be
\Cp = \{\Om, \OL \text{ or } w, h \}
\ee 
where $\Om$ is the matter density (in units of the critical energy density), $\OL$ is the dark energy density, $w$ is the dark energy equation of state (taken to be redshift-independent, although this assumption can easily be generalized) and $h$ is defined as $H_0 = 100 h$km/s/Mpc, where $H_0$ is the value of the Hubble constant today\footnote{The Hubble parameter $h$ actually plays the role of a nuisance parameter, as it cannot be constrained by distance modulus measurements independently unless the absolute magnitude of the SNe is known, for the two quantities are perfectly degenerate.}. The curvature parameter $\Omk$ is related to the matter and dark energy densities by the constraint equation
\be
\Omk = 1 - \Om - \OL.
\ee
In the following, we shall consider either a Universe with non-zero curvature (with $\Omk \neq 0$ and an appropriate prior) but where the dark energy is assumed to be a cosmological constant, i.e.\ with $w=-1$ (the \lcdm{} model), or a flat Universe where the effective equation of state parameter is allowed to depart from the cosmological constant value, i.e. $\Omk = 0, w \neq -1$ (the wCDM model).

In a Friedman-Robertson-Walker cosmology defined by the parameters $\Cp$, the distance modulus to a SN at redshift $z_i$ is given by 
\be \label{eq:mucosmo}
\mu_i = \mu(z_i, \Cp) = 5 \log \left[ \frac{D_L(z_i, \Cp)}{\Mpc} \right] +25,
\ee
where $D_L$ denotes the luminosity distance to the SN. This can be rewritten as 
\be \label{eq:theory_distance_modulus}
\mu_i = \eta + 5\log d_L(z_i,\Om, \OL, w),
\ee
where
\be \label{eq:defeta}
\eta = -5\log\frac{100h}{c} + 25
\ee
and $c$ is the speed of light in km/s. We have defined the dimensionless luminosity distance (with $D_L = c/H_0 d_L$, where $c$ is the speed of light)
\begin{equation}
d_L(z,\Om, \OL, w) = \frac{(1+z)}{\sqrt{|\Omk|}}\text{sinn} \{ \sqrt{|\Omk|}  \int_0^z \dr z' 
\[ (1+z')^3\Om + \Ode(z') + (1+z')^2\Omk \]^{-1/2} \} 
\end{equation}
with the dark energy density parameter 
\be
\Ode(z) =  \OL \exp\left(3\int_0^z \frac{1+w(x)}{1+x}\dr x \right).
\ee
In the above equation, we have been completely general about the functional form of the dark energy equation of state, $w(z)$. In the rest of this work, however, we will make the further assumption that $w$ is constant with redshift, i.e., $w(z) = w$. We have defined the function $\text{sinn}(x) = x, \sin(x), \sinh(x)$ for a flat Universe ($\Omk = 0$), a closed Universe ($\Omk < 0$) or an open Universe, respectively.

The problem is now to infer, given data $\data$ in Eq.~\eqref{eq:data}, the values (and uncertainties) of the cosmological parameters $\Cp$, as well as the nuisance parameters  $\{\alpha, \beta \}$, appearing in Eq.~\eqref{eq:observed_distance_modulus} and any further parameter describing the SNe population and its intrinsic scatter. Before building a full Bayesian hierarchical model to solve this problem, we briefly describe the usual approach and its shortcomings.

\subsection{Shortcomings of the usual $\chisq$ method} \label{sec:chisq}

The usual analysis (e.g., \cite{Astier:2005qq,Kowalski2008Improved,Kessler2009Firstyear,Conley:2011ku}) defines a $\chisq$ statistics as follows: 
\be \label{eq:chisq}
\chisq_\mu = \sum_i \frac{(\mu_i - \mu_i^\text{obs})^2}{\sigma_{\mu i}^2}.
\ee
where $\mu_i$ is given by Eq.~\eqref{eq:theory_distance_modulus} as a function of the cosmological parameters and the ``observed'' distance modulus $\mu_i^\text{obs}$ is obtained by replacing in Eq.~\eqref{eq:observed_distance_modulus} the best-fit values for the colour and stretch correction and B-band magnitude from the \salt{} output (denoted by hats). Furthermore, the intrinsic magnitude for each SN, $M_i$, is replaced by a global parameter $M$, which represents the mean intrinsic magnitude of all SNe in the sample:
\be \label{eq:mu_relaced}
\mu_i^\text{obs} =  \hat{m}_{Bi}^* - M + \alpha \cdot \hat{x}_{1i} - \beta \cdot \hat{c}_i \, ,
\ee
where the mean intrinsic magnitude $M$ is unknown. The variance $\sigma^2_{\mu i}$ comprises several sources of uncertainty, which are added in quadrature:
\be
\sigma_{\mu i}^2 = (\sigma_{\mu i}^\text{fit})^2 + (\sigma_{\mu i}^z)^2 + (\sigint)^2,
\ee
where $\sigma_{\mu i}^\text{fit}$ is the statistical uncertainty from the \salt{} lightcurve fit, 
\be
\left(\sigma_{\mu i}^\text{fit}\right)^2 = \ul{\Psi}^T \Cm_i \ul{\Psi}
\ee
where $\Psi = \left(1, \alpha, -\beta \right)$ and $\Cm_i $ is the covariance matrix given in Eq.~\eqref{eq:def_Cm}. $\sigma_{\mu i}^z$ is the uncertainty on the SN redshift from spectroscopic measurements and peculiar velocities of and within the host galaxy; finally, $\sigint$ is an unknown parameter describing the SN intrinsic dispersion. Further discussions of the unknown $\sigint$ estimation problem, see \citep{BlondinMandel2011,Kim2011,VishwakarmaNarlikar2011}. As  mentioned above, ideally $\sigint$ is a single quantity that encapsulates the remaining intrinsic dispersion in the SNIa sample, folding in all of the residual scatter due to physical effects not well captured by the Phillips corrections. However, observational uncertainties such as the estimation of photometric errors can lead to a variation of $\sigint$ sample by sample (for which there is a growing body of evidence). While we do not consider the latter scenario in this paper, it is important to keep in mind that describing the whole SN population with a single scatter parameter $\sigint$ is likely to be an oversimplification.   

Further error terms are added in quadrature to the total variance, describing uncertainties arising from dispersion due to lensing, Milky Way dust extinction, photometric zero-point calibration, etc. In this work, we do not deal with such systematic uncertainties, though they can be included in our method and we comment further on this below.

The cosmological parameter fit proceeds by minimizing the $\chi^2$ in Eq.~\eqref{eq:chisq}, simultaneously fitting the cosmological parameters, $\alpha$, $\beta$ and the mean intrinsic SN magnitude $M$.  The value of $\sigint$ is adjusted to obtain $\chisq_\mu/\text{dof} \sim 1$ (usually on a sample-by-sample basis), often rejecting individual SNe with a residual pull larger than some arbitrarily chosen cut-off. It was recognized early that fitting the numerator and denominator of Eq.~\eqref{eq:chisq} iteratively leads to a large ``bias'' in the recovered value of $\beta$~\citep{Kowalski2008Improved,Astier:2005qq,Wang:2005bw}. This has been traced back to the fact that the error on the colour correction parameter $c_i$ is as large as or larger than the width of the distribution of values of $c_i$, especially for high-redshift SNe. This is a crucial observation, which constitutes the cornerstone of our Bayesian hierarchical model, as explained below. We demonstrate that an appropriate modeling of the distribution of values of $c_i$ leads to an effective likelihood that replaces the $\chi^2$ of Eq.~\eqref{eq:chisq}. With this effective likelihood and appropriate Bayesian priors, all parameters can be recovered without bias. If instead one adopts a properly normalized likelihood function, i.e., replacing the $\chi^2$ of Eq.~\eqref{eq:chisq} with
\be
\like = L_0 \exp\left(-\frac{1}{2}{\chi_\mu^2} \right)
\ee
(with the pre-factor $L_0$ chosen so that the likelihood function integrates to unity in data space), marginalization over $\alpha, \beta$ leads to catastrophic biases in the recovered values (up to $\sim 6\sigma$ in some numerical tests we performed). This is a strong hint that the naive form of the likelihood function above is incorrect for this problem. The effective likelihood we derive below solves this problem.

The standard approach to cosmological parameters fitting outlined above adopted in most of the literature to date has several shortcomings, which can be summarized as follows: 
\begin{enumerate}

\item The expression for the $\chisq$, Eq.~\eqref{eq:chisq}, has no fundamental statistical justification, but is based on a heuristic derivation. The fundamental problem with Eq.~\eqref{eq:chisq} is that some of the parameters being fitted (namely, $\alpha, \beta$) control both the location and the dispersion of the $\chisq$ expression, as they appear both in the numerator and the denominator, via the $(\sigma_{\mu i}^\text{fit})^2$ term. Therefore, the statistical problem is one of jointly estimating both the location and the variance. We show below how this can be tackled using a principled statistical approach. 

\item Adjusting $\sigint$ to obtain the desired goodness-of-fit is problematic, as it does not allow one to carry out any further goodness-of-fit test on the model itself, for obviously the variance has been adjusted to achieve a good fit by construction. This means that model checking is by construction not possible with this form of the likelihood function.

\item It would be interesting to obtain not just a point estimate for $\sigint$, but an actual probability distribution for it, as well. This would allow consistency checks e.g. among different surveys, to verify whether the recovered intrinsic dispersions are mutually compatible (within errorbars). This is currently not possible given the standard $\chisq$ method. A more easily generalizable approach is desirable, that would allow one to test the hypothesis of multiple SNe populations with different values of intrinsic dispersion, for example as a consequence of evolution with redshift, or correlated with host galaxy properties. Current practice is to split the full SN sample in subsamples (e.g., low- and high-redshift, or for different values of the colour correction) and check for the consistency of the recovered values from each of the subsamples. Our method allows for a more systematic approach to this kind of important model checking procedure.

\item It is common in the literature to obtain inferences on the parameters of interest by minimizing (i.e., profiling) over nuisance parameters entering in Eq.~\eqref{eq:chisq}. This is in general much more computationally costly than marginalization from e.g. MCMC samples~\citep{Feroz:2011bj}. There are also examples where some nuisance parameters are marginalized over, while others are maximised~\citep{Astier:2005qq}, which is statistically inconsistent and should best be avoided. It should also be noted that maximisation and marginalization do not in general yield the same errors on the parameters of interest when the distribution is non-Gaussian. From a computational perspective, it would be advantageous to adopt a fully Bayesian method, which can be used in conjunction with fast and efficient MCMC and nested sampling techniques for the exploration of the parameter space. This would also allow one to adopt Bayesian model selection methods (which cannot currently be used with the standard $\chisq$ approach as they require the full marginalization of parameters to compute the Bayesian evidence).

\item The treatment of systematic errors is being given great attention in the recent literature (see e.g.~\cite{Nordin:2008aa}), but the impact of various systematics on the final inference for the interesting parameters, $\Cp$, has often been propagated in an approximate way (e.g., \cite{Kessler2009Firstyear}, Appendix F). As we are entering an epoch when SN cosmology is likely to be increasingly dominated by systematic errors, it would be desirable to have a consistent way to include sources of systematic uncertainties in the cosmological fit and to propagate the associated error consistently on the cosmological parameters. The inclusion in the analysis pipeline of systematic error parameters has been hampered so far by the fact that this increases the number of parameters being fitted above the limit of what current methods can handle. However, if a fully Bayesian expression for the likelihood function was available, one could then draw on the considerable power of Bayesian methods (such as MCMC, or nested sampling) which can efficiently handle larger parameter spaces. 
\end{enumerate}

Motivated by the above problems and limitations of the current standard method, we now proceed to develop in the next section a fully Bayesian formalism from first principles, leading to the formulation of a new effective likelihood which will overcome, as will be shown below, the above problems. A more intuitive understanding of our procedure can be acquired from the simpler toy problem described in Appendix~\ref{app:toy}.

%%%%%%%%%%%%%%%%%%%%%%%%%%%%%%%%%%%%%%%%%%%%%%%%%
%%%%%%%%%%%%% EFFECTIVE LIKELIHOOD FOR SNIa %%%%%%%%%%%%%%%% 
%%%%%%%%%%%%%%%%%%%%%%%%%%%%%%%%%%%%%%%%%%%%%%%%%

\subsection{The Bayesian hierarchical model} \label{sec:bhm}

\begin{figure}
\centering
\includegraphics[width=0.6 \linewidth]{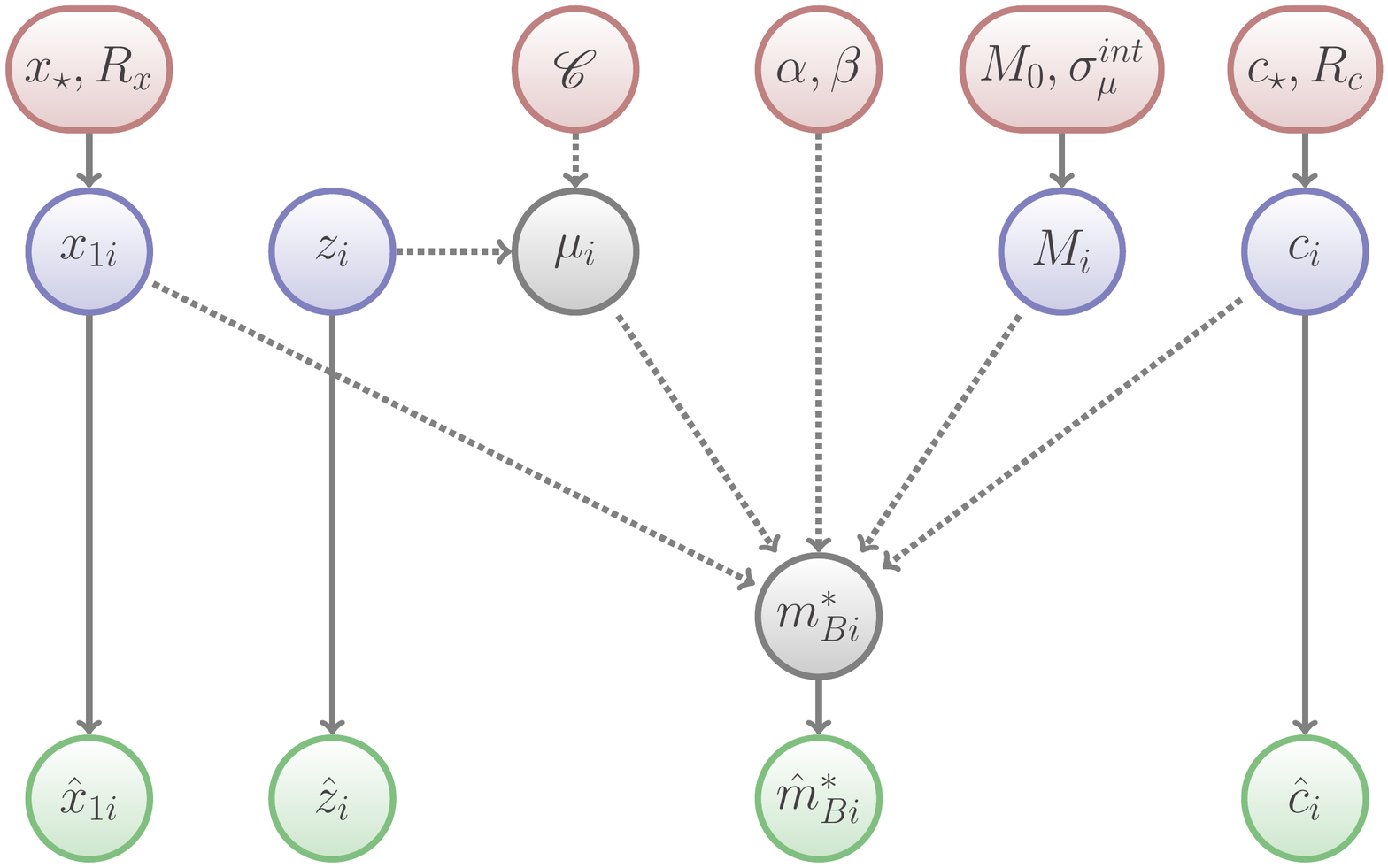}
\caption{Graphical network showing the deterministic (dashed) and probabilistic (solid) connections between variables in our Bayesian hierarchical model (BHM). Variables of interest are in red, latent (unobserved) variables are in blue and observed data (denoted by hats) are in green.}
\label{fig:sn_network}
\end{figure}
We now turn to developing a Bayesian hierarchical model (BHM) for the SNe data from \salt{} lightcurve fits. The same general linear regression problem with unknown variance has been addressed by~\cite{Kelly:2007jy}, and applied in that paper to X-ray spectral slope fitting. 
The gist of our method is shown in the graphical network of Fig.~\ref{fig:sn_network}, which displays the probabilistic and deterministic connection between variables. The fundamental idea is that we introduce a new layer of so-called ``latent'' variables -- that is, quantities which describe the ``true'' value of the corresponding variables, and which are obviously unobserved (represented in blue in Fig.~\ref{fig:sn_network}). In particular, we associate with each SN a set of latent variables  $(z_i, c_i, x_{1i}, M_i)$, which give the true value of the redshift, colour correction, stretch correction and intrinsic magnitude for that SN. For the stretch and colour correction, we have noisy estimates in the form of data $\hat{c}_i, \hat{x}_{1i}$ (with an associated covariance matrix) resulting from the \salt{} fits (solid vertical lines). For the redshift $z_i$, we have noisy estimates given by the measured values $\hat{z}_i$. As above, all observed quantities are denoted by a hat.

Each of the latent variables $(c_i, x_{1i}, M_i)$ is modeled as an uncorrelated random variable drawn from an underlying parent distribution, described by a Gaussian with two parameters, a population mean and a variance\footnote{In this paper we use the notation:
\be
x \sim \Nd(m,\sigma^2)
\ee
to denote a random variable $x$ being drawn from an underlying Gaussian distribution with mean $m$ and variance $\sigma^2$. In vector notation, $m$ is replaced by a vector $\underline{m}$, while $\sigma^2$ is replaced by the covariance matrix $\Sigma$ and the distribution is understood to be the appropriate multidimensional Normal, see Appendix~\ref{sec:notation}. }. The latent intrinsic magnitudes $M_i$ are thought of as realizations from a SN population with overall mean intrinsic magnitude $M_0$ and intrinsic dispersion $\sigint$ (both to be estimated from the data):
\be \label{eq:Mi_normal}
M_i \sim \Nd(M_0, (\sigint)^2).
\ee
This is represented in Fig.~\ref{fig:sn_network} by the solid arrow connecting the parameters $M_0, (\sigint)^2$ with the latent variables $M_i$. This represents statistically the physical statement that an intrinsic dispersion $(\sigint)^2$ is expected to remain in the distribution of SNe magnitudes, even after applying the Phillips corrections to the observed magnitudes. The simple {\em Ansatz} of Eq.~\eqref{eq:Mi_normal} can be replaced by a more refined modeling, which describes the existence of multiple SN populations, for example with intrinsic magnitude correlated with host galaxy properties, for which there is a growing body of evidence~\citep{Sullivan:2006ah,Mandel:2010xj,Sullivan:2011kv}. We shall investigate this possibility in a forthcoming work. 

The parent distribution of the true colour and stretch corrections is similarly represented by Gaussians, again parameterised each by a mean ($\cnot, \xnot$) and a variance ($\Rc^2, \Rs^2$) as 
\be \label{eq:ci_x1i_normal}
c_i \sim \Nd( \cnot,\Rc^2), \quad x_{1i} \sim \Nd(\xnot, \Rs^2).
\ee
As above, the quantities $ \cnot,\Rc^2, \xnot, \Rs^2$ have to be estimated from the data. The choice of a Gaussian distribution for the latent variables $\cbl$ and $\xbl$ is justified by the fact that the observed distribution of $\cbh$ and $\xbh$, shown in Fig.~\ref{fig:c_x1_hist} for the actual SNIa sample described in section~\ref{sec:data} below, is fairly well described by a Gaussian. As shown in Fig.~\ref{fig:c_x1_hist}, there might be a hint for a heavier tail for positive values of $\cbh$, but this does not fundamentally invalidate our Gaussian approximation. It would be easy to expand our method to consider other distributions, for example mixture models of Gaussians to describe a more complex population or a distribution with heavier tails, if non-Gaussianities in the observed distribution should make such modeling necessary. In this paper we consider the simple uni-modal Gaussians given by Eq.~\eqref{eq:ci_x1i_normal}.  

%-----------------------FIGURE REAL c x1 HISTOGRAMS  --------------------------------------
\begin{figure*}
\centering
\includegraphics[width=0.8\linewidth]{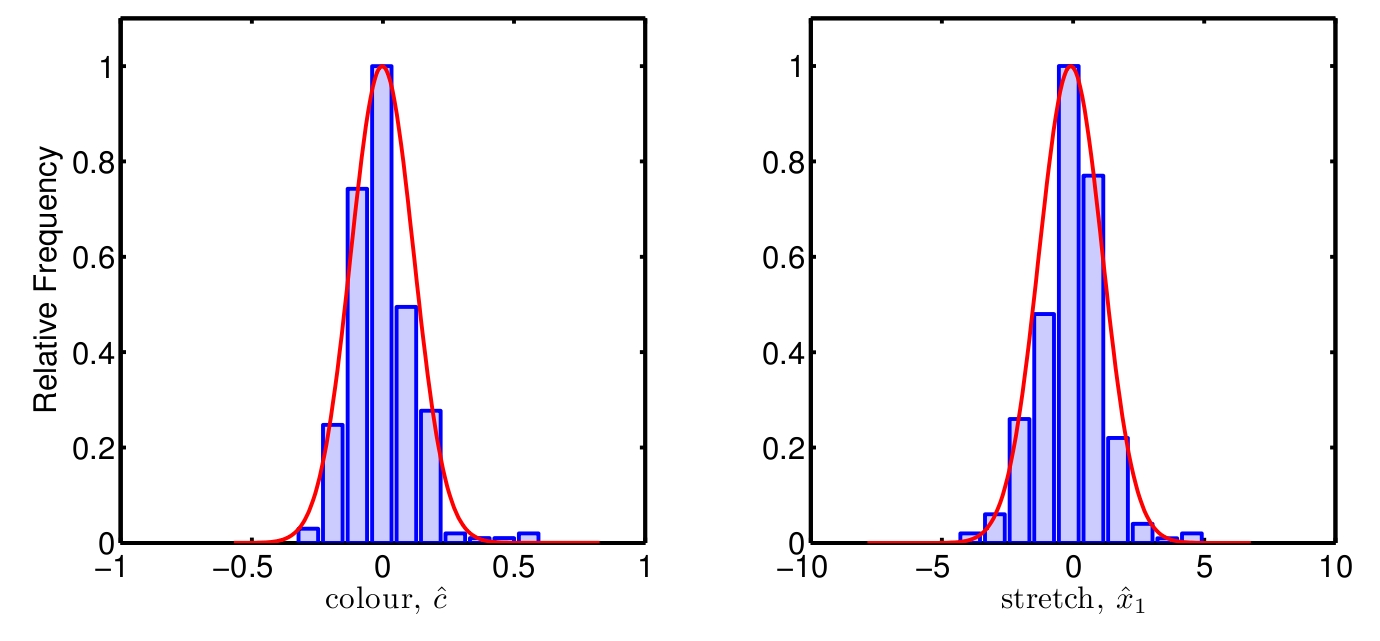}
\caption{Histogram of observed stretch parameters  $\hat{x}_{1i}$ and observed colour parameters $\hat{c}_{i}$ from the 288 SNIa from Kessler et al. (2009a), compared with a Gaussian fit (red curve).}
\label{fig:c_x1_hist}
\end{figure*}
%-----------------------FIGURE  REAL c x1 HISTOGRAMS --------------------------------------

While we are interested in the value (and distribution) of the SNe population parameters $M_0, (\sigint)^2$, the parameters entering Eqs.~\eqref{eq:ci_x1i_normal} are not physically interesting. Therefore they will be marginalized out at the end. It is however crucial to introduce them explicitly in this step: it is precisely the lack of modeling in the distribution of $c_i$ that usually leads to the observed ``biases'' in the reconstruction of $\beta$, see Appendix~\ref{app:toy} for an explicit example with a toy model. For ease of notation, we re-write Eq.~\eqref{eq:Mi_normal} and \eqref{eq:ci_x1i_normal} 
in matrix notation as:
\begin{align} 
\mg & \sim \Nd(\mg_0,\Zd), \label{eq:m_normal_matrix}\\
\cbl & \sim \Nd(\cnot \cdot \onesn, \diag{\Rc^2 \cdot \onesn}) =  p(\cbl|\cnot,\Rc)  \label{eq:c_normal_matrix}\\
\xbl & \sim \Nd(\xnot \cdot \onesn, \diag{\Rs^2 \cdot \onesn}) = p(\xbl|\xnot,\Rs) \label{eq:x_normal_matrix}
\end{align}
where 
\begin{align}
\mg & = (M_1, \dots, M_n)  \in \Real^n, \\
\mg_0 &  = M_0 \cdot \onesn  \in \Real^n, \\
\Zd & = \diag{(\sigint)^2\cdot \onesn} \in \Real^{n\times n}.
\end{align}

Having introduced $3n$ latent (unobserved) variables ($\cbl, \xbl, \mg$), where $n$ is the number of SNe in the sample, the fundamental strategy of our method is to link them to underlying population parameters via Eqs.~\eqref{eq:Mi_normal} and \eqref{eq:ci_x1i_normal}, then to use the observed noisy estimates to infer constraints on the population parameters of interest (alongside the cosmological parameters), while marginalizing out the unobserved latent variables. 

The intrinsic magnitude $M_i$ is related to the observed $B$-band magnitude $\hat{m}_B^*$ and the distance modulus $\mu$ by Eq.~\eqref{eq:observed_distance_modulus}, which can be rewritten in vector notation as:
\be \label{eq:phil2vec}
\mbl = \underline{\mu}+  \mg -\alpha \xbl +\beta \cbl.
\ee
Note that the above relation is {\em exact}, i.e. $\mg,  \xbl, \cbl$ are here the latent variables (not the observed quantities), while $\mbl$ is the true value of the $B$-band magnitude (also unobserved). This is represented by the dotted (deterministic) arrows connecting the variables in Fig.~\ref{fig:sn_network}.

We seek to determine the posterior pdf for the parameters of interest $\Theta = \{ \Cp, \alpha, \beta, \sigint \}$, while marginalizing over the uknown population mean intrinsic magnitude, $M_0$. From Bayes theorem, the marginal posterior for $\Theta$ is given by (see e.g.~\cite{Trotta:2008qt,BMIC:2010} for applications of Bayesian methods to cosmology):
\be
p( \Theta | \data ) = \int {\rm d} M_0 p( \Theta, M_0 | \data ) =  \int {\rm d} M_0 \frac{p(\data | \Theta, M_0 )p(\Theta, M_0)}{p(D)},
\label{eq:posterior}
\ee
where $p(D)$ is the Bayesian evidence (a normalizing constant) and the prior $p(\Theta, M_0)$ can be written as
\be
p(\Theta, M_0) = p(\Cp, \alpha, \beta)p(M_0, \sigint) = p(\Cp, \alpha, \beta) p(M_0|\sigint)p(\sigint).
\label{eq:prior_theta}
\ee
We take a uniform prior on the variables $\Cp, \alpha, \beta$ (on a sufficiently large range so as to encompass the support of the likelihood), as well as a Gaussian prior for $p(M_0|\sigint)$, since $M_0$ is a location parameter of a Gaussian (conditional on $\sigint$). Thus we take 
\be
p(M_0|\sigint) =  \norm_{M_0}(M_m, \sigma^2_{M_0}),
\ee
where the mean of the prior ($M_m = -19.3$ mag) is taken to be a plausible value in the correct ballpark, and the variance ($\sigma_{M_0} = 2.0$ mag) is sufficiently large so that the prior is very diffuse and non-informative (the precise choice of mean and variance for this prior does not impact on our numerical results). Finally, the appropriate prior for $\sigint$ is a Jeffreys' prior, i.e., uniform in $\log\sigint$, as $\sigint$ is a scale parameter, see e.g.~\cite{Box:1992}. Although the intrinsic magnitude $M_0$ and the Hubble constant $H_0$ are perfectly degenerate as far as SNIa data are concerned, we do not bundle them together in a single parameter but treat them separately with distinct priors, as we are interested in separating out the variability due to the distribution of the SNIa intrinsic magnitude. 
 
We now proceed to manipulate further the likelihood, $p(\data | \Theta, M_0 ) = p(\cbh,\xbh,\mbh | \Theta, M_0)$:

\begin{align}
p(\cbh,\xbh,\mbh|\Theta, M_0)&= \int \dr \cbl \, \,\dr \xbl \, \, \dr \mg \, \,p(\cbh,\xbh,\mbh|\cbl,\xbl,\mg,\Theta, M_0) p(\cbl,\xbl,\mg|\Theta, M_0) \\
&= \int \dr \cbl \, \,\dr \xbl \, \, \dr \mg \, \,p(\cbh,\xbh,\mbh|\cbl,\xbl,\mg,\Theta)  \notag \\
& \times  \int\dr \Rc \, \, \dr \Rs \,\, \dr \cnot \,\, \dr \xnot \,\, \,\,p(\cbl | \cnot,\Rc) p(\xbl | \xnot,\Rs) p(\mg|M_0, \sigint)
  p(\Rc)p(\Rs)p(\cnot)p(\xnot) %\\ 
%&=\int \dr \cbl \, \,\dr \xbl \, \, \dr \mg \, \,p(\cbh,\xbh,\mbh|\cbl,\xbl,\mg,\Theta, M_0) p(\cbl,\xbl|\Theta, M_0) p(\mg|\Theta, M_0) \\
%&=\int \dr \cbl \, \,\dr \xbl \, \, \dr \mg \, \,p(\cbh,\xbh|\mbh,\cbl,\xbl,\mg,\Theta, M_0) p(\mbh|\cbl,\xbl,\mg,\Theta, M_0)  p(\cbl,\xbl|\Theta, M_0) p(\mg|\Theta, M_0) . 
\label{eq:eff_like_1}
\end{align}
In the first line, we have introduced a set of $3n$ latent variables, $\{ \cbl,\xbl,\mg \}$, which describe the {\em true} value of the colour, stretch and intrinsic magnitude for each SNIa. Clearly, as those variables are unobserved, we need to marginalize over them. In the second line, we have replaced $p(\cbl,\xbl,\mg|\Theta, M_0)$ by the distributions of the latent $\{ \cbl,\xbl,\mg \}$ given by the probabilistic relationships of Eq.~\eqref{eq:m_normal_matrix} and Eqs.~(\ref{eq:c_normal_matrix}--\ref{eq:x_normal_matrix}), obtained by marginalizing out the population parameters $\{ \Rc, \Rs, \cnot, \xnot \}$:
\be
p(\cbl,\xbl, \mg |\Theta, M_0) = \int \dr \Rc \,\,\dr \Rs \,\,\dr \cnot \,\,\dr \xnot \,\, p(\cbl|\cnot,\Rc) p(\xbl|\xnot,\Rs) p(\mg|M_0, \sigint) p(\Rc)p(\Rs)p(\cnot)p(\xnot).
\ee
(we have also dropped $M_0$ from the likelihood, as conditioning on $M_0$ is irrelevant if the latent $\mg$ are given). If we further marginalize over $M_0$ (as in Eq.~\eqref{eq:posterior}, including the prior on $M_0$), the expression for the effective likelihood, Eq.~\eqref{eq:eff_like_1}, then becomes: 
\begin{align}
p(\cbh,\xbh,\mbh|\Theta)&=
 \int \dr \cbl \, \,\dr \xbl \, \, \dr \mg \, \,p(\cbh,\xbh,\mbh|\cbl,\xbl,\mg,\Theta)  \notag \\
& \times  \int\dr \Rc \, \, \dr \Rs \,\, \dr \cnot \,\, \dr \xnot \,\, \dr M_0 \,\,p(\cbl | \cnot,\Rc) p(\xbl | \xnot,\Rs) p(\mg|M_0, \sigint)
  p(\Rc)p(\Rs)p(\cnot)p(\xnot)p(M_0 | \sigint)
\label{eq:eff_like_2}
\end{align}

The term $p(\cbh,\xbh,\mbh|\cbl,\xbl,\mg,\Theta)$ is the conditional probability of observing values $\{ \cbh,\xbh,\mbh \} $ if the latent (true) value of $\cbl,\xbl,\mg$ and of the other cosmological parameters were known. From Fig.~\ref{fig:sn_network}, $\mbl$ is connected only deterministically to all other variables and parameters, via Eq.~\eqref{eq:phil2vec}. Thus we can replace $\mbl = \muu + \mg - \alpha \cdot \xbl + \beta \cdot \cbl$ and write  
\begin{align}\label{eq:mobs}
p(\cbh,\xbh,\mbh|\cbl,\xbl,\mg,\Theta) & = \prod_{i=1}^n \norm(\mu_i + M_i - \alpha \cdot x_{1i} + \beta \cdot c_i, \Cm_i) \\
& = |2 \pi \Scov|^{-\hf} \exp \left( - \hf [(X-X_0)^T\Scov^{-1}(X-X_0)] \right)
\end{align}
where $\mu_i \equiv \mu_i(z_i, \Theta)$ and we have defined 
\begin{align}
X&= \{X_1, \dots,X_n \} \in \Real^{3n}, \quad X_0=\{ X_{0,1}, \dots,X_{0,n}\} \in \Real^{3n}, \\
X_i&= \{ c_i, x_{1,i},(M_i-\alpha x_{1,i}+\beta c_i)\} \in \Real^{3}, \quad X_{0,i}=\{ c_i,x_{1,i},\mhi-\mu_i\} \in \Real^{3}, 
\end{align}
as well as the $3n\times 3n$ block covariance matrix\footnote{Notice that we neglect correlations between different SNIa, which is reflected in the fact that $\Sigma_C$ takes a block-diagonal form. It would be however very easy to add arbitrary cross-correlations to our formalism (e.g., coming from correlated systematic within survey, for example zero point calibration) by adding such non-block diagonal correlations to Eq.~\eqref{eq:Sigma_C}.} 
\be \label{eq:Sigma_C}
\Sigma_C = \left( \begin{array}{cccc}
\Cm_1 & 0 & 0 & 0  \\
0 & \Cm_2 & 0 & 0\\
0 & 0 & \ddots & 0 \\
0 & 0 & 0 & \Cm_n \end{array} \right) \, .
\ee

Finally we explicitly include redshift uncertainties in our formalism. The observed apparent magnitude, $\mbh$, on the left-hand-side of Eq.~\eqref{eq:mobs}, is the value at the observed redshift, $\zbh$. However, $\muu$ in Eq.~\eqref{eq:mobs} should be evaluated at the true (unknown) redshift, $\zbl$. As above, the redshift uncertainty is included by introducing the latent variables $\zbl$ and integrating over them:
\begin{align}\label{eq:z_integral}
p(\cbl,\xbl, \mg |\cbl,\xbl,\mg,\Theta) &= \int \dr \zbl \,\,\,\, p(\cbl,\xbl, \mg |\cbl,\xbl,\mg,\zbl,\Theta)p(\zbl|\zbh)
\end{align}
where we model the redshift errors $p(\zbl|\zbh)$ as Gaussians:
\be
\zbl \sim \norm(\zbh, \Zz)
\ee
with a $n \times n$ covariance matrix:
\be
\Zz = \text{diag}(\sigma_{z_1}^2, \dots, \sigma_{z_n}^2).
\ee
It is now necessary to integrate out all latent variables and nuisance parameters from the expression for the likelihood, Eq.~\eqref{eq:eff_like_2}. This can be done analytically, as all necessary integral are Gaussian. The detailed steps are described in Appendix~\ref{app:like}. In summary, the procedure consists of:
\begin{enumerate}
\item Marginalization over the intrinsic redshifts, see Eq.~\eqref{eq:marginalized_z}.
\item Marginalization over the latent variables $\{\mg,\cbl, \xbl\}$, see Eq.~\eqref{eq:eff_nlike_2}.
\item Marginalization over the nuisance parameters $\{ \xnot ,\cnot \}$, see Eq.~\eqref{eq:eff_nlike_3}. 
\end{enumerate}
This leads to the final expression for the effective likelihood of Eq.~\eqref{eq:eff_like_final}:
\be
\begin{aligned}
p(\cbh,\xbh,\mbh|\Theta)&= \int \dr \log \Rc \,\,\dr \log \Rs \,\,  |2 \pi\Scov|^{-\hf} |2\pi \Sigp|^{-\hf} |2\pi \Sa|^{\hf} |2\pi \So|^{-\hf} |2\pi K|^{\hf} \notag \\
& \times  \exp \left( - \hf [X_0^T\Scov^{-1}X_0 -\Delta^T\Sa \Delta -k_0^TK^{-1}k_0 +\bub_m^T \So^{-1}\bub_m] \right), 
 \label{eq:eff_like_final_text}
\end{aligned}
\ee
where the various vectors and matrices appearing in the above expression are defined in Appendix~\ref{app:like}. This equation is the main result of our paper. The two remaining nuisance parameters $\Rc, \Rs$ cannot be integrated out analytically, so they need to be marginalized numerically. Hence, they are added to our parameters of interest and are sampled over numerically, and then marginalized out from the joint posterior.

%
%%%%%%%%%%%%%%%%%%%%%%%%%%%%%%%%%%%%%%%%%%%%%%%%%
%%%%%%%%%%%%% NUMERICAL SIMULATIONS %%%%%%%%%%%%%%%% 
%%%%%%%%%%%%%%%%%%%%%%%%%%%%%%%%%%%%%%%%%%%%%%%%%

%************************************************************************************
\section{Numerical tests of the method}
\label{sec:tests}
%************************************************************************************
\subsection{Generation of simulated SNe data} 
\label{sec:data}

We now proceed to test the performance of our method on simulated data sets, and to compare it with the usual $\chi^2$ approach. We take as starting point the recent compilation of 288 SNIa from~\cite{Kessler2009Firstyear}, the data set on which we apply our Bayesian hierarchical model in the next section. \cite{Kessler2009Firstyear} re-fitted the lightcurves from five different surveys:
\begin{itemize}
\item SDSS: 103 SNe~\citep{Kessler2009Firstyear},
\item ESSENCE: 56 SNe \citep{MiknaitisPignata2007,WoodVassey2007},
\item SNLS: 62 SNe \citep{Astier:2005qq},
\item Nearby Sample: 33 SNe \citep{JhaRiess2007} and
\item HST: 34 SNe \citep{GarnavichKirshner1998,KnopAldering2003,RiessStrolger2004,RiessStrolger2007},
\end{itemize}
using both the \salt{} method and the \mlcs{} fitter. In the following, we are exclusively employing the results of their \salt{} fits and use those as the observed data set for the purposes of our current work, as described in the previous section. More refined procedures could be adopted, for example by simulating lightcurves from scratch, using e.g.~the publicly available package SNANA~\citep{Kessler:2009yy}. In this paper we chose a simpler approach, consisting of simulating \salt{} fit results in such a way to broadly match the distributions and characteristics of the real data set used in \cite{Kessler2009Firstyear}. 

%--------/\------True params and priors ------/\------
\begin{table}
\centering
\begin{tabular}{lll}
\hline\hline 
Parameter & Symbol & True Value   \\ \hline 
Matter energy density parameter&$\Om$ & 0.3  \\
Dark energy density parameter &$\OL$ & 0.7  \\
Dark energy equation of state & $w$ & $-1$ \\
Spatial curvature & $\Omk$ & 0.0 \\ 
Hubble expansion rate &$H_0$ [km/s/Mpc] & 72.0 \\\hline
Mean absolute magnitude of SNe & $M_0$ [mag]  & -19.3   \\
Intrinsic dispersion of SNe magnitude& $\sigint$ [mag] & 0.1 \\ 
Stretch correction & $\alpha$& 0.13 \\
Colour correction  & $\beta$ & 2.56 \\ \hline
Mean of distribution of  $\xbl$ & $\xnot$ & 0.0 \\
Mean of distribution of  $\cbl$ & $\cnot$ & 0.0 \\
s.d. of distribution of  $\xbl$ & $\Rs$ & 1.0 \\
s.d. of distribution of $\cbl$ & $\Rc$ & 0.1 \\
Observational noise on $\mb$ & $\sigma_{\mbi}$ & Depending on survey \\
Observational noise on $\xbl$ & $\sigma_{x_1i}$ & Depending on survey \\
Observational noise on $\cbl$ & $\sigma_{ci}$ & Depending on survey \\
Correlation between $\xbl$ and $\cbl$ & $\sigma_{x_1i,ci}$ & 0.0 \\
\hline
%51 & 2 & 3 & 4 & 5 & 6 \\ \hline
\end{tabular}
\caption{Input parameter values used in the generation of the simulated SNe \salt{} fits.}
\label{tab:simsn}
\end{table}

The numerical values of the parameters used in the simulation are shown in Table \ref{tab:simsn}. We adopt a vanilla, flat $\Lambda$CDM cosmological model as fiducial cosmology. The Phillips correction parameters are chosen to match the best-fit values reported in \cite{Kessler2009Firstyear}, while the distributional properties of the colour and stretch correction match the observed distribution of their total SN sample. For each survey, we generate a number of SNe matching the observed sample, and we model their redshift distribution as a Gaussian, with mean and variance estimated from the observed distribution within each survey. The observational error of $\mb, \cbl, \xbl$ is again drawn from a Gaussian distribution whose mean and variance have been matched to the observed ones for each survey. Finally, the pseudo-data (i.e., the simulated \salt{} fits results) are generated by drawing from the appropriate distributions centered around the latent variables. For simplicity, we have set to 0 the off-diagonal elements in the correlation matrix~\eqref{eq:Sigma_C} in our simulated data, and neglected redshift errors. None of these assumptions have a significant impact on our results. In summary, our procedure for each survey is as follows:
\begin{enumerate}
\item Draw a value for the latent redshift $z_i$ from a normal distribution with mean and variance matching the observed ones. As we neglect redshift errors in the pseudo-data for simplicity (since the uncertainty in $z$ is subdominant in the overall error budget), we set $\hat{z}_i = z_i$.
\item Compute $\mu_i$ using the fiducial values for the cosmological parameters $\Cp$ and the above $z_i$ from Eq.~\eqref{eq:mucosmo}.
\item Draw the latent parameters $x_{1i}, c_i, M_i$ from their respective distributions (in particular, including an intrinsic scatter $\sigint = 0.1$ mag in the generation of $M_i$).
\item Compute $\mbi$ using $x_{1i}, c_i, M_i$ and the Phillips relation Eq.~\eqref{eq:observed_distance_modulus}.
%\item Draw a redshift standard deviation $\sigma_{z_i}$ from a normal distribution matching the observed distribution in the survey.
\item Draw the value of the standard deviations $\sigma_{x_1i}, \sigma_{c_i},\sigma_{m_i},$ from the appropriate normal distributions for each survey type. A small, $z_i$-dependent stochastic linear addition is also made to $\sigma_{x_1i}, \sigma_{c_i},\sigma_{m_i},$ to mimic the observed correlation between redshift and error.
%\item Draw the observed redshift $\hat{z}_i$ from $\hat{z}_i \sim \norm (z_i, \sigma_{z_i})$. \rt{what do you use for $\sigma_{z_i}$?}
\item Draw the \salt{} fit results from  $\hat{x}_{1i} \sim \norm (x_{1i}, \sigma_{x_1i})$, $\hat{c}_{i} \sim \norm (c_{i}, \sigma_{c_i})$ and $\mbih \sim \norm (\mbi, \sigma_{m_i})$.
\end{enumerate}    
%-----------------------FIGURE SIM DATA  --------------------------------------
\begin{figure*}
\centering
\includegraphics[width=0.8\linewidth]{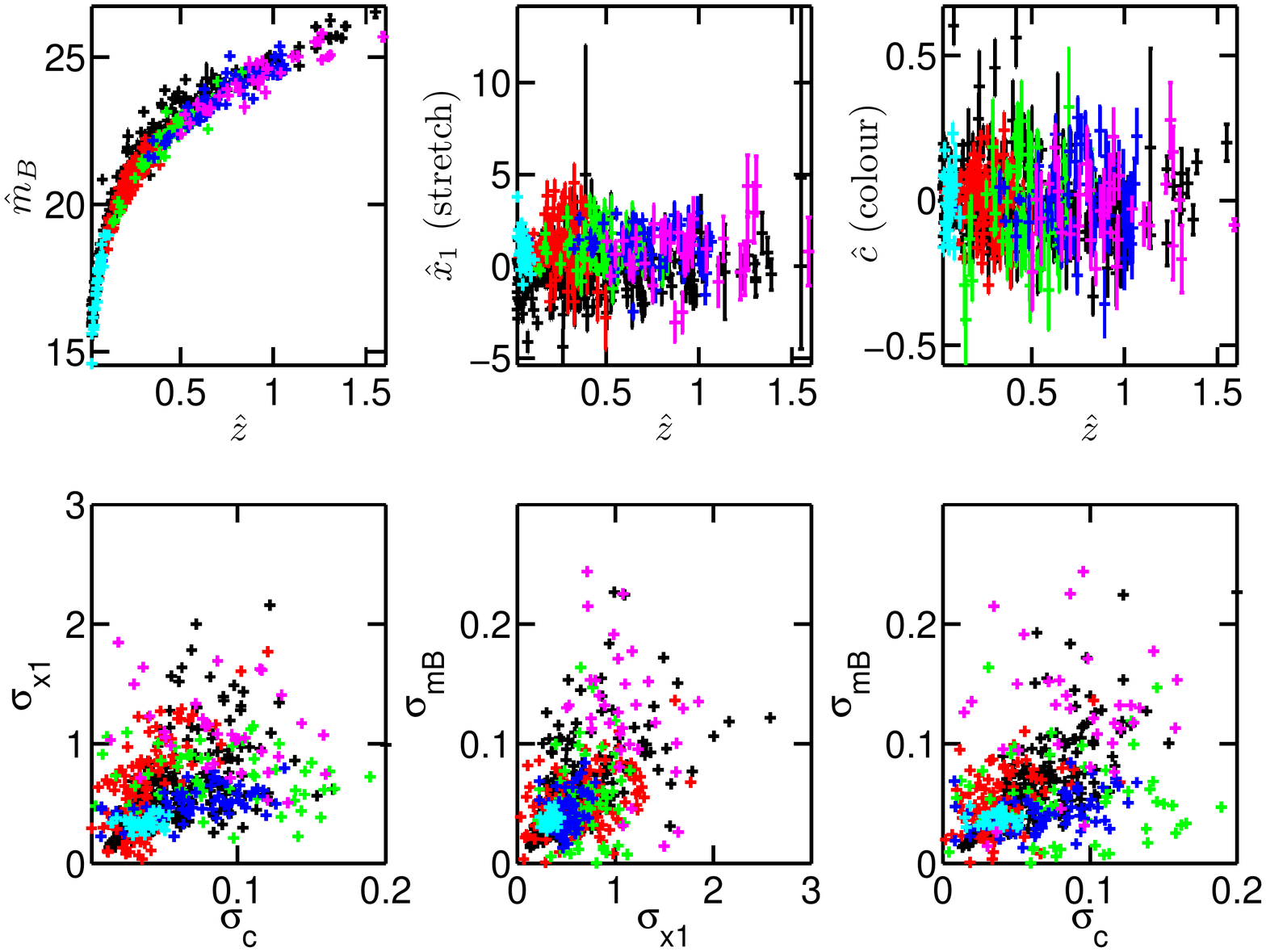}
\caption{An example realization of our simulated data sets (coloured according to survey), superimposed on real data (black). Colour code for simulated data survey: nearby sample (cyan), ESSENCE (green), SDSS (red), SNLS (blue) and HST (magenta).}
\label{fig:mixdathist}
\end{figure*}
%-----------------------FIGURE SIM DATA  --------------------------------------

As shown in Fig.~\ref{fig:mixdathist}, the simulated data from our procedure have broadly similar distributions to the to real ones. The two notable exceptions are the overall vertical shift observed in the distance modulus plot, and the fact that our synthetic data cannot reproduce the few outliers with large values of the variances (bottom panels). The former is a consequence of the different intrinsic magnitude used in our simulated data (as the true one is unknown). However, the intrinsic magnitude is marginalized over at the end, so this difference has no impact on our inferences. The absence of outliers is a consequence of the fact that our simulation is a pure phenomenological description of the data, hence it cannot encapsulate such fine details. While in principle we could perform outlier detection with dedicated Bayesian procedures, we do not pursue this issue further in this paper. We stress once more that the purpose of our simulations is not to obtain realistic SNIa data. Instead, they should only provide us with useful mock data sets coming from a known model so that we can test our procedure. More sophisticated tests based on more realistically generated data (e.g., from SNANA) are left for future work. 

\begin{table*} %--------\/------priors------\/------
\begin{tabular}{l l l }
\hline 
Parameter        &  \lcdm{}                  & wCDM\\ \hline \hline
$\Om$            & Uniform: $\mc{U}(0.0,1.0)$ & Uniform: $\mc{U}(0.0,1.0)$ \\
$\Omk$           &Uniform: $\mc{U}(-1.0,1.0)$& Fixed: $  0     $ \\
$w$              & Fixed: $-1$                           & Uniform: $\mc{U}(-4,0)$   \\
$H_0$ [km/s/Mpc] & $\norm(72, 8^2)$ & $\norm(72, 8^2)$\\ \hline
 & \multicolumn{2}{c}{Common priors} \\\hline 
$\sigint$ [mag]  & \multicolumn{2}{c}{Uniform on $\log \sigint$: $\mc{U}(-3.0,0.0)$}\\
$M_0$ [mag]      &  \multicolumn{2}{c}{Uniform: $\mc{U}(-20.3,-18.3) $}\\
$\alpha$         &  \multicolumn{2}{c}{Uniform: $\mc{U}(0.0, 1.0) $}\\
$\beta$          &  \multicolumn{2}{c}{Uniform: $\mc{U}(0.0, 4.0) $}\\
$\Rc$            &  \multicolumn{2}{c}{Uniform on $\log \Rc $: $\mc{U}(-5.0,2.0)$}\\
$\Rs$            &  \multicolumn{2}{c}{Uniform on $\log \Rc $: $\mc{U}(-5.0, 2.0)$}\\

\hline  
\end{tabular}
\caption{Priors on our model's parameters used when evaluating the posterior distribution. Ranges for the uniform priors have been chosen so as to generously bracket plausible values of the corresponding quantities.} 
\label{tab:priors}
\end{table*}   %--------/\------priors ------/\------
%

%

%************************************************************************************
\subsection{Numerical sampling}

After analytical marginalization of the latent variables, we are left with the following 8 parameters entering the effective likelihood of Eq.~\eqref{eq:eff_like_final_text}: 
\be \label{eq:params}
\{ \Om, \Omk \text{ or } w, H_0, \sigint, \alpha, \beta, \Rc, \Rs \} \, .
\ee
As mentioned above, in keeping with the literature we only consider either flat Universes with a possible $w\neq -1$ (the $\Lambda$CDM model), or curved Universes with a cosmological constant ($w=-1$, the wCDM model). Of course it is possible to relax those assumptions and consider more complicated cosmologies with a larger number of free parameters if one so wishes (notably including evolution in the dark energy equation of state). 

Of the parameters listed in Eq.~\eqref{eq:params},  the quantities $\Rc, \Rs$ are of no interest and will be marginalized over. As for the remaining parameters, we are interested in building their marginal 1 and 2-dimensional posterior distributions. This is done by plugging the likelihood \eqref{eq:eff_like_final_text} into the posterior of Eq.~\eqref{eq:posterior}, with priors on the parameters chosen according to Table~\ref{tab:priors}. We use a Gaussian prior on the Hubble parameter $H_0 = 72 \pm 8$ km/s/Mpc from local determinations of the Hubble constant~\citep{Freedman:2000cf}. However, as $H_0$ is degenerate with the intrinsic population absolute magnitude $M_0$ (which is marginalized over at the end), replacing this Gaussian prior with a less informative prior $H_0 \text{[km/s/Mpc]} \sim {\mc U}(20,100)$ has no influence on our results. 

Numerical sampling of the posterior is carried out via a nested sampling algorithm~\citep{Skilling2004,Skilling2006,Feroz2008,Feroz2009}. 
Although the original motivation for nested sampling was to compute 
the Bayesian evidence, the recent development of the 
MultiNest algorithm \citep{Feroz2008,Feroz2009} has delivered an 
extremely powerful and versatile algorithm that has been demonstrated 
to be able to deal with
extremely complex likelihood surfaces in hundreds of dimensions
exhibiting multiple peaks. 
As samples from the posterior are
generated as a by-product of the evidence computation, nested
sampling can also be used to obtain parameter constraints in the
same run as computing the Bayesian evidence. In this paper we adopt the publicly available 
MultiNest algorithm~\citep{Feroz2008} to obtain samples from the posterior distribution of Eq.~\eqref{eq:posterior}. We use 4000 live points and a tolerance parameter 0.1, resulting in about $8\times 10^5$ likelihood evaluations.\footnote{A Fortran code implementing our method is available from the authors upon request.}

We also wish to compare the performance of our BHM with the usually adopted $\chi^2$ minimization procedure. To this end, we fit the pseudo-data using the $\chi^2$ expression of Eq.~\eqref{eq:chisq}. In order to mimic what is done in the literature as closely as possible, we first fix a value of $\sigint$. Then, we simultaneously minimize the $\chi^2$ w.r.t. the fit parameters $\vartheta = \{ \Om, \Omk \text{ or } w, H_0, M_0, \alpha, \beta \}$, as described below. We then evaluate the $\chi^2/\text{dof}$ from the resulting best fit point, and we adjust $\sigint$ to obtain  $\chi^2/\text{dof} = 1$. We then repeat the above minimization over $\vartheta$ for this new value of $\sigint$, and iterate the procedure. Once we have obtained the global best fit point, we derive 1- and 2-dimensional confidence intervals on the parameters by profiling (i.e., maximising over the other parameters) over the likelihood  
\be
L(\vartheta) = \exp\left(-\frac{1}{2}\chi(\vartheta)^2\right),
\ee
with $\chi^2$  given by Eq.~\eqref{eq:chisq}. According to Wilks' theorem, approximate confidence intervals are obtained from the profile likelihood as the regions where the $\chi^2$ increases by $\Delta\chi^2$ from its minimum value, where $\Delta\chi^2$ can be computed from the chi-square distribution with the number of degree of freedoms corresponding to the number of parameters of interest and is given in standard look-up tables.  

Obtaining reliable estimates of the profile likelihood using Bayesian algorithms (such as MultiNest) is a considerably harder numerical task than mapping out the Bayesian posterior. However, it has been shown that MultiNest can be successfully used for this task even in highly challenging situations~\citep{Feroz:2011bj}, provided the number of live points and tolerance value used are adjusted appropriately. For our $\chi^2$ scan, we adopt $10^4$ live points and a tolerance of 0.1. We have found that those values give accurate estimates of the profile likelihood more than $2\sigma$ into the tails of the distribution for an 8 dimensional Gaussian toy model (whose dimensionality matches the case of interest here). With these MultiNest settings, we gather $1.5 \times 10^5$ samples, from which the profile likelihood is derived. 

Our implementation of the $\chi^2$ method is designed to match the main features of the fitting procedure usually adopted in the literature (namely, maximisation of the likelihood rather than marginalization of the posterior, and iterative determination of the intrinsic dispersion), although we do not expect that it exactly reproduces the results obtained by any specific implementation. Its main purpose is to offer a useful benchmark against which to compare the performance of our new Bayesian methodology.

\subsection{Parameter reconstruction}

%-----------------------FIGURE SIM DATA CONTOUR PLOTS --------------------------------------
\begin{figure*}
\centering
\includegraphics[width=0.48\linewidth]{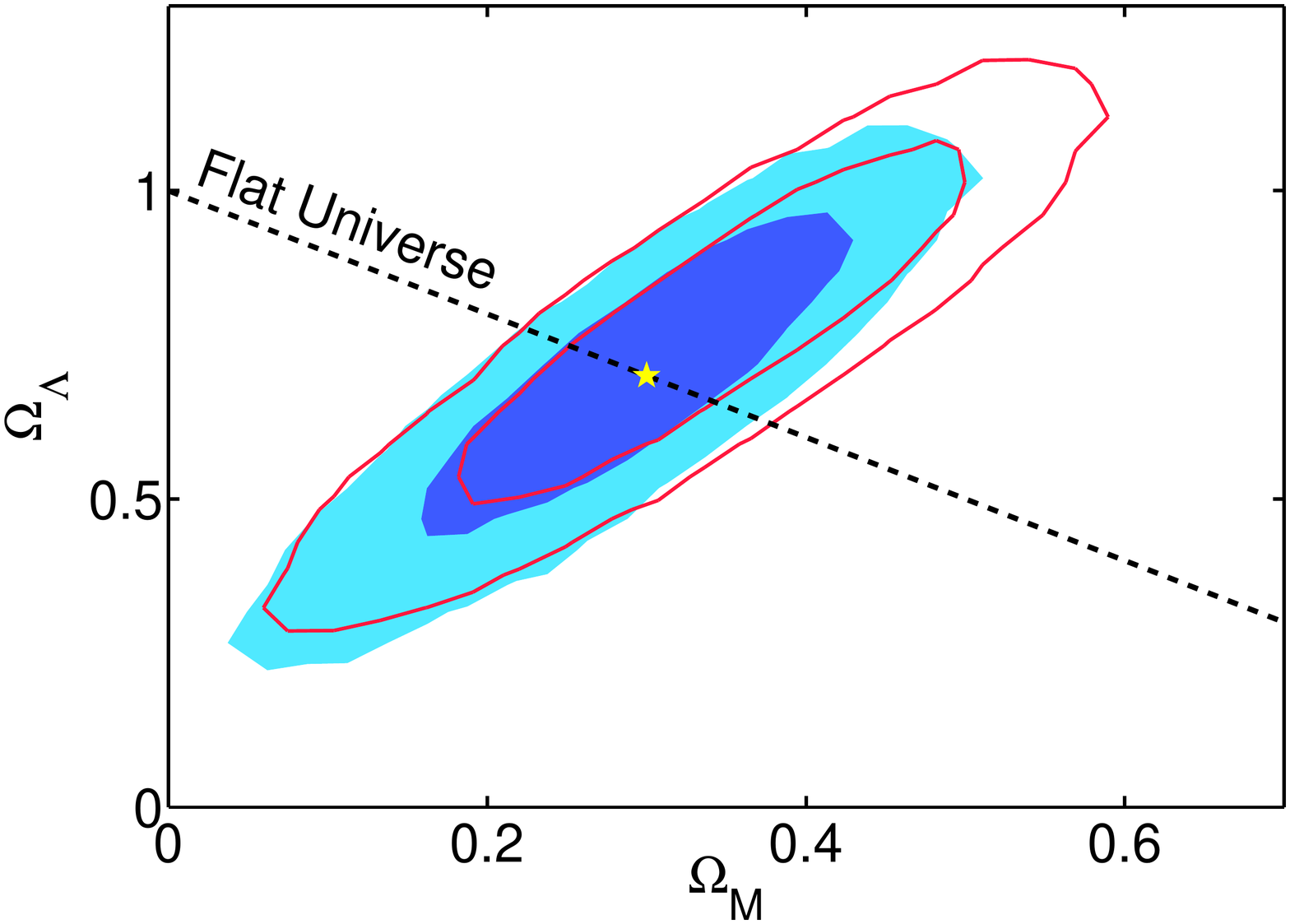}\hfill
\includegraphics[width=0.48\linewidth]{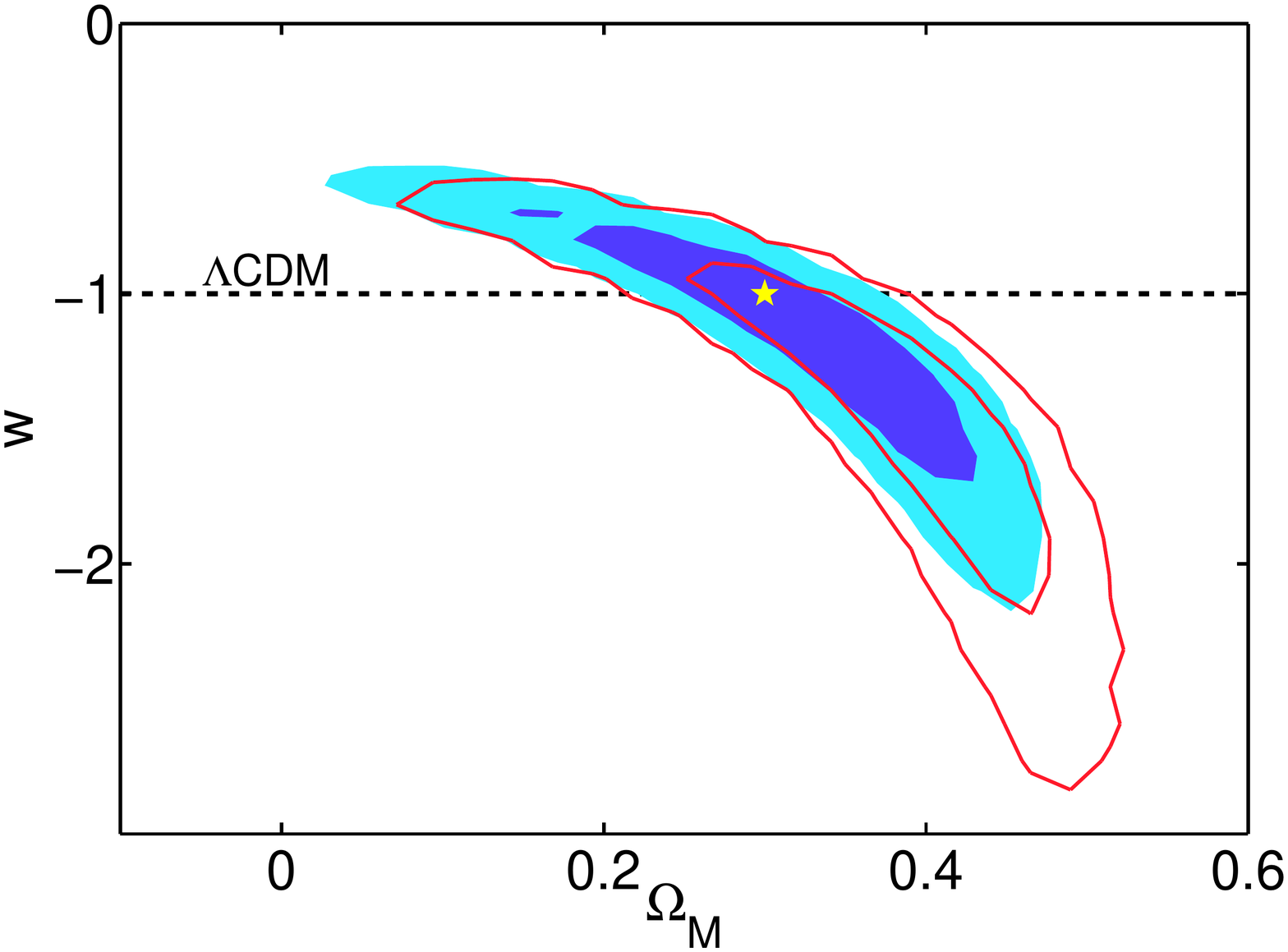}
\caption{Reconstruction of cosmological parameters from a simulated data set encompassing 288 SNIa, with characteristics matching presently available surveys (including realization noise). Blue regions contain 95\% and 68\% of the posterior probability (other parameters marginalized over) from our BHM method, the red contours delimit 95\% and 68\% confidence intervals from the standard $\chi^2$ method (other parameters maximised). The yellow star indicates the true value of the parameters. The left panel assumes $w=-1$ while the right panel assumes $\Omk = 0$. Notice how out method produced considerably less biassed constraints on the parameters. }
\label{fig:compare}
\end{figure*}
%-----------------------FIGURE SIM DATA CONTOUR PLOTS --------------------------------------

We compare the cosmological parameters reconstructed from the standard $\chisq$ method and our Bayesian approach in Fig.~\ref{fig:compare} for a typical data realization. The left-hand-side panel shows constraints in the $\Om-\OL$ plane for the $\Lambda$CDM model, both from our Bayesian method (filled regions, marginalized 68\% and 95\% posterior) and from the standard $\chi^2$ method (red contours, 68\% and 95\% confidence regions from the profile likelihood). In the right-hand-side panel, constraints are shown in the $w-\Om$ plane instead (this fit assumes a flat Universe). In a typical reconstruction, our Bayesian method produced considerably tighter constraints on the cosmological parameters of interest than the usual $\chi^2$ approach. Our constraints are also less biassed w.r.t. the true value of the parameters, an important advantage that we further characterize below. 

\begin{figure*}
\centering
\hfill
\includegraphics[width=0.33\linewidth]{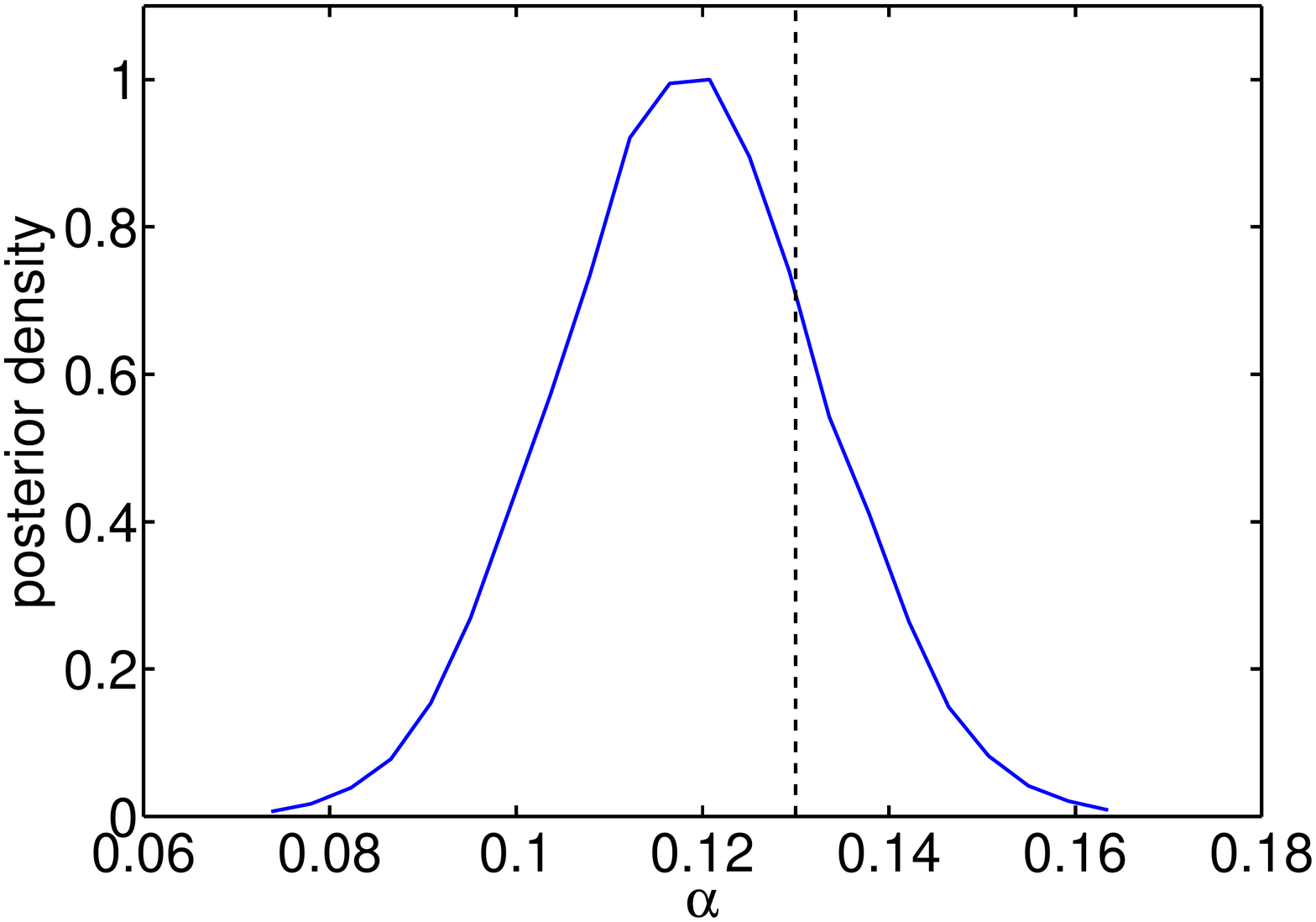} \hfill
\includegraphics[width=0.33\linewidth]{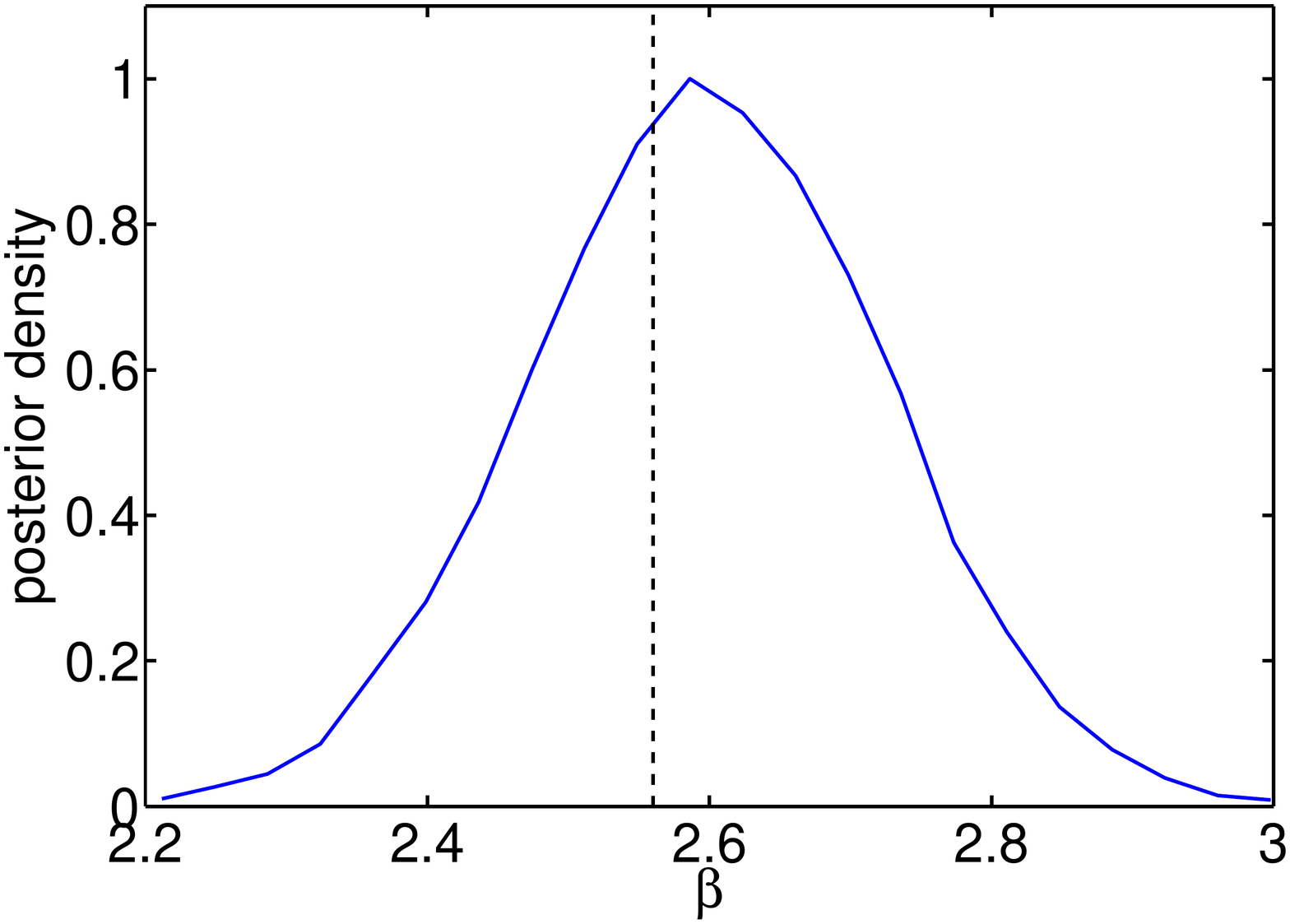} \hfill
\includegraphics[width=0.33\linewidth]{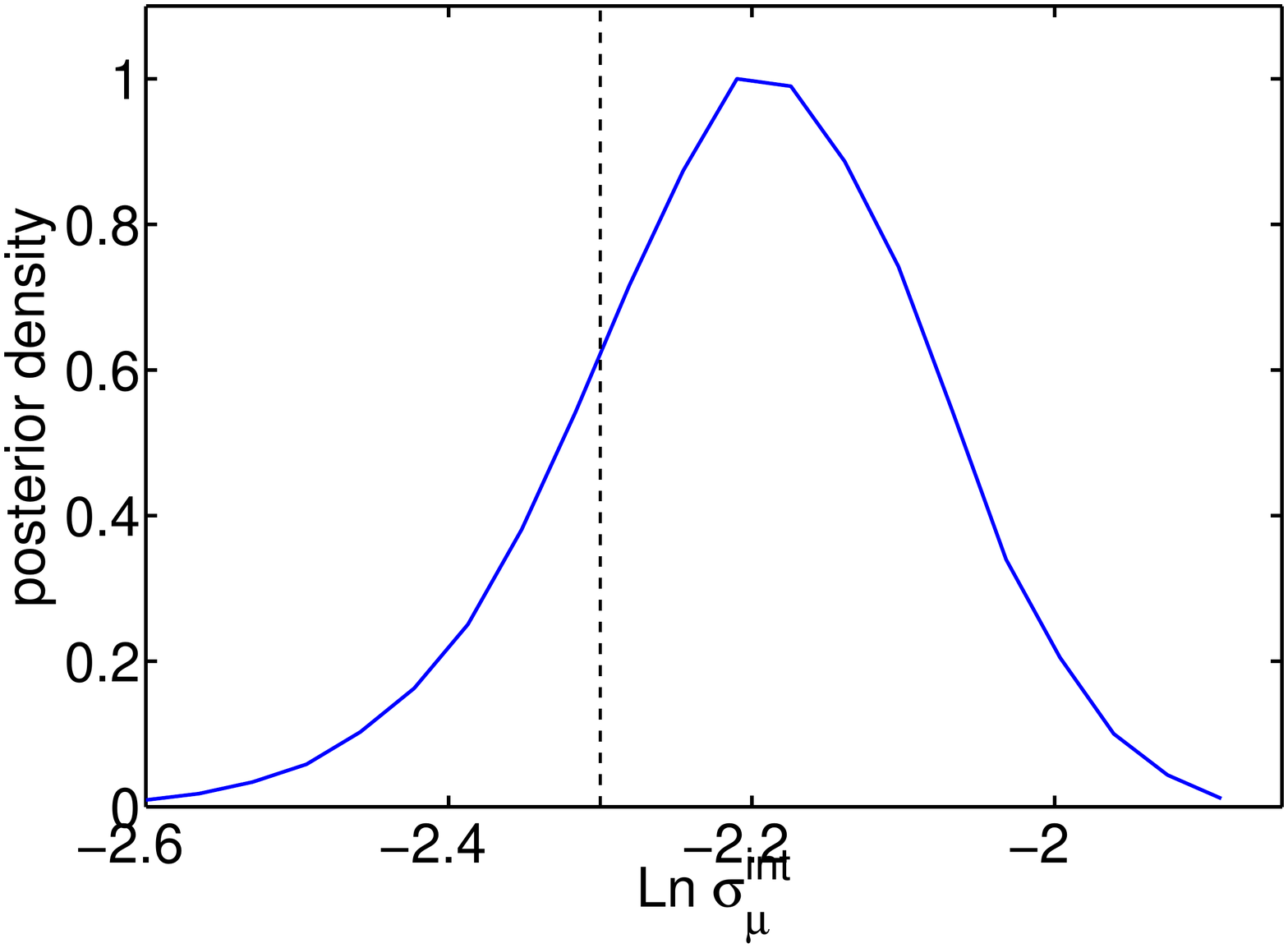} 
\caption{Marginalised posterior for the stretch correction $\alpha$, colour correction $\beta$ parameter and logarithm of the intrinsic dispersion of SNe, $\log\sigint$, from a simulated data set from our Bayesian method. The vertical, dashed line gives the true value for each quantity.}
\label{fig:1d_posteriors}
\end{figure*}

Our BHM further produces marginalized posterior distributions for all the other parameters of the fit, including the stretch and colour corrections and the intrinsic dispersion of the SNe. The 1D marginal posteriors for those quantities are shown in Fig.~\ref{fig:1d_posteriors}.  The recovered posterior means lie within $1\sigma$ of the true values. Notice that we do not expect the posterior mean to match exactly the true value, because of realization noise in the pseudo-data. However, as shown below, our method delivers less biassed estimates of the parameters, and a reduced mean squared error compared with the standard $\chisq$ approach. The stretch correction $\alpha$ is determined with $8\%$ accuracy, while the colour correction parameter $\beta$ is constrained with an accuracy better than $3\%$. A new feature of our method is that it produces a posterior distribution for the SN population intrinsic dispersion, $\sigint$ (right-hand-side panel of Fig~\ref{fig:1d_posteriors}). This allows to determined the intrinsic dispersion of the SNIa population to typically about 10\% accuracy.

%
%-----------------------FIGURE SIM Sig Int --------------------------------------
%-----------------------FIGURE SIM Sig Int --------------------------------------

%
%
\subsection{Comparison of long-term performance of the two methods}
Encouraged by the good performance of our BHM on single instances of simulated data, we now wish to compare the long-term performance of the two methods for a series of simulated data realizations. We are interested in comparing the average ability of both methods to recover parameter values that are as much as possible unbiased with respect to their true values, as well as to establish the coverage properties of the credible and confidence intervals. 

Coverage is defined as the probability that an interval contains (covers) the true value of a parameter, in a long series of repeated measurements. The defining property of a e.g.~95\% frequentist confidence interval is that it should cover the true value 95\% of the time; thus, it is reasonable to check if the intervals have the properties they claim.  Coverage is a frequentist concept: intervals based on Bayesian techniques are meant to contain a given amount of posterior probability for a {\em single measurement} (with no reference to repeated measurements) and are referred to as credible intervals to emphasize the difference in concept.  While Bayesian techniques are not designed with coverage as a goal, it is still meaningful to investigate their coverage properties. To our knowledge, the coverage properties of even the standard $\chisq$ method (which, being a frequentist method would ideally be expected to exhibit exact coverage) have never been investigated in the SN literature. 

\begin{figure*}
\centering
\includegraphics[width=0.49 \linewidth]{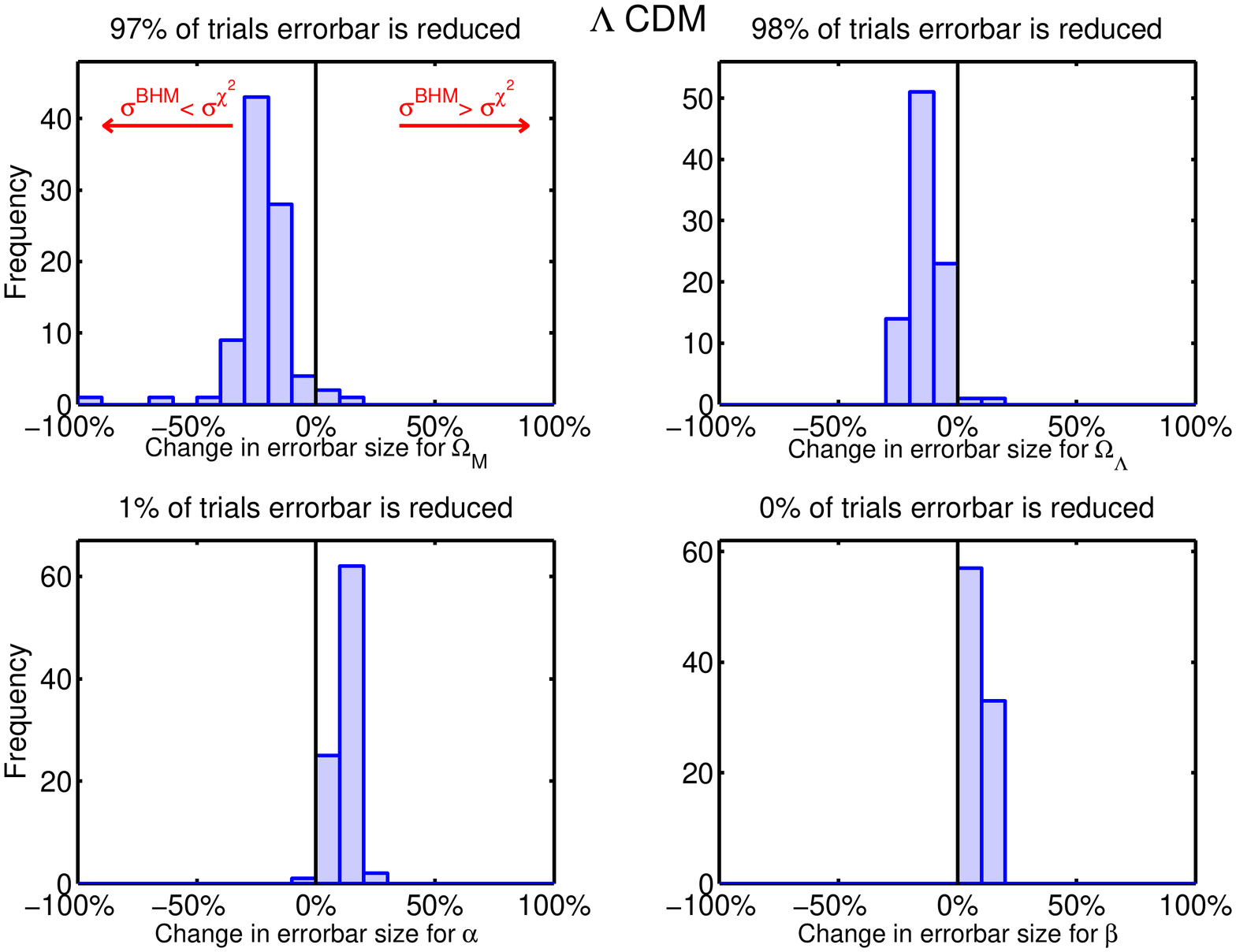} \hfill
\includegraphics[width=0.49 \linewidth]{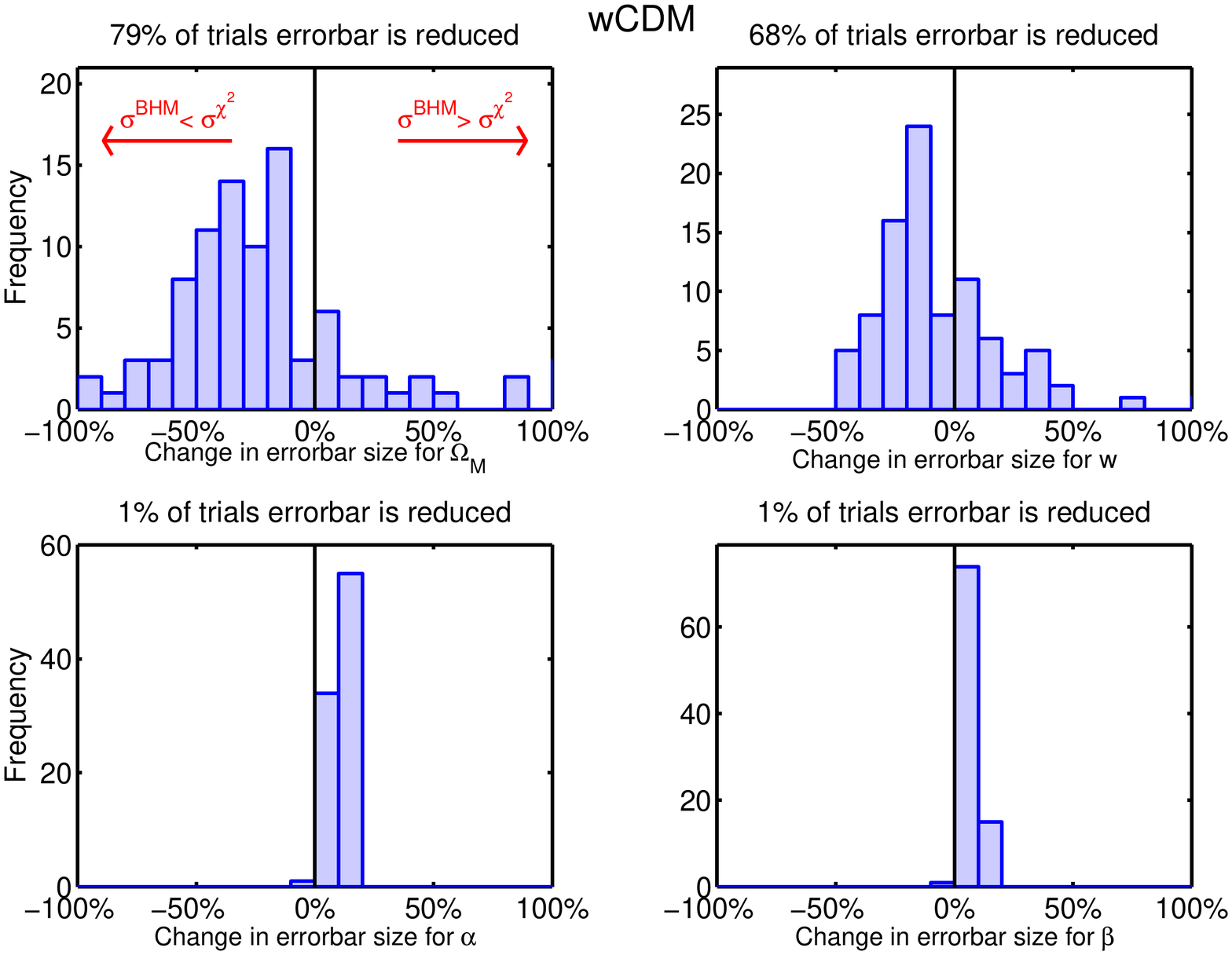} 
\caption{Histograms of the quantity defined in Eq.~\eqref{eq:Stheta}, comparing of the errorbars on each parameter from our method and from the standard $\chi^2$ approach for 100 realization, for the $\Lambda$CDM model (left) and the wCDM model (right). A change in errorbar size of $-10\%$ indicates BHM errorbars are $10\%$ smaller than $\chi^2$ errorbars.  A change in errorbar size of $+10\%$ indicates BHM errorbars are $10\%$ larger than $\chi^2$ errorbars. Our BHM method generally delivers smaller errors on the cosmological parameters (top row) but larger errors on the Phillips correction parameters (bottom row).}
\label{fig:mix7a}
\end{figure*}

We generate 100 realizations of the pseudo-data from the fiducial model of Table~\ref{tab:simsn} as described in section~\ref{sec:data}, and we analyze them using our BHM method and the standard $\chisq$ approach, using the same priors as above, given in Table~\ref{tab:priors}. For each parameter of interest $\theta$, we begin by considering the relative size of the posterior 68.3\% range from out method, $\sigma_\theta^\text{BHM}$, compared with the 68.3\% confidence interval from the $\chi^2$ method, $\sigma_\theta^{\chi^2}$, which is summarized by the quantity ${\mathcal S}_\theta$ which shows the percentage change in errorbar size with respect to the errorbar derived using the $\chi^2$ method
%\be \label{eq:Stheta} %ORIGINAL DEFINITION PRIOR TO 29.07.2011
%{\mathcal S}_\theta \equiv 1 - \frac{\sigma_\theta^{\chi^2}}{\sigma_\theta^\text{BHM}}.
%\ee 
\be \label{eq:Stheta} %NEW DEFINITION  29.07.2011
{\mathcal S}_\theta \equiv  \left( \frac{\sigma_\theta^\text{BHM}}{\sigma_\theta^{\chi^2}}-1 \right) \times 100.
\ee 
A value ${\mathcal S}_\theta  < 1$ means that our method delivers tighter errorbars on the parameter $\theta$. A histogram of this quantity for the variables of interest is shown in Fig.~\ref{fig:mix7a}, from which we conclude that our method gives smaller errobars on $\Om, \OL$ and $w$ in almost all cases. The flip side of this result is that the uncertainty on $\alpha, \beta$ is larger from our method than from the $\chi^2$ approach in essentially all data realization. This is a consequence of the large number of latent variables our method marginalizes over.

\begin{figure*}
\centering
\includegraphics[width=0.49 \linewidth]{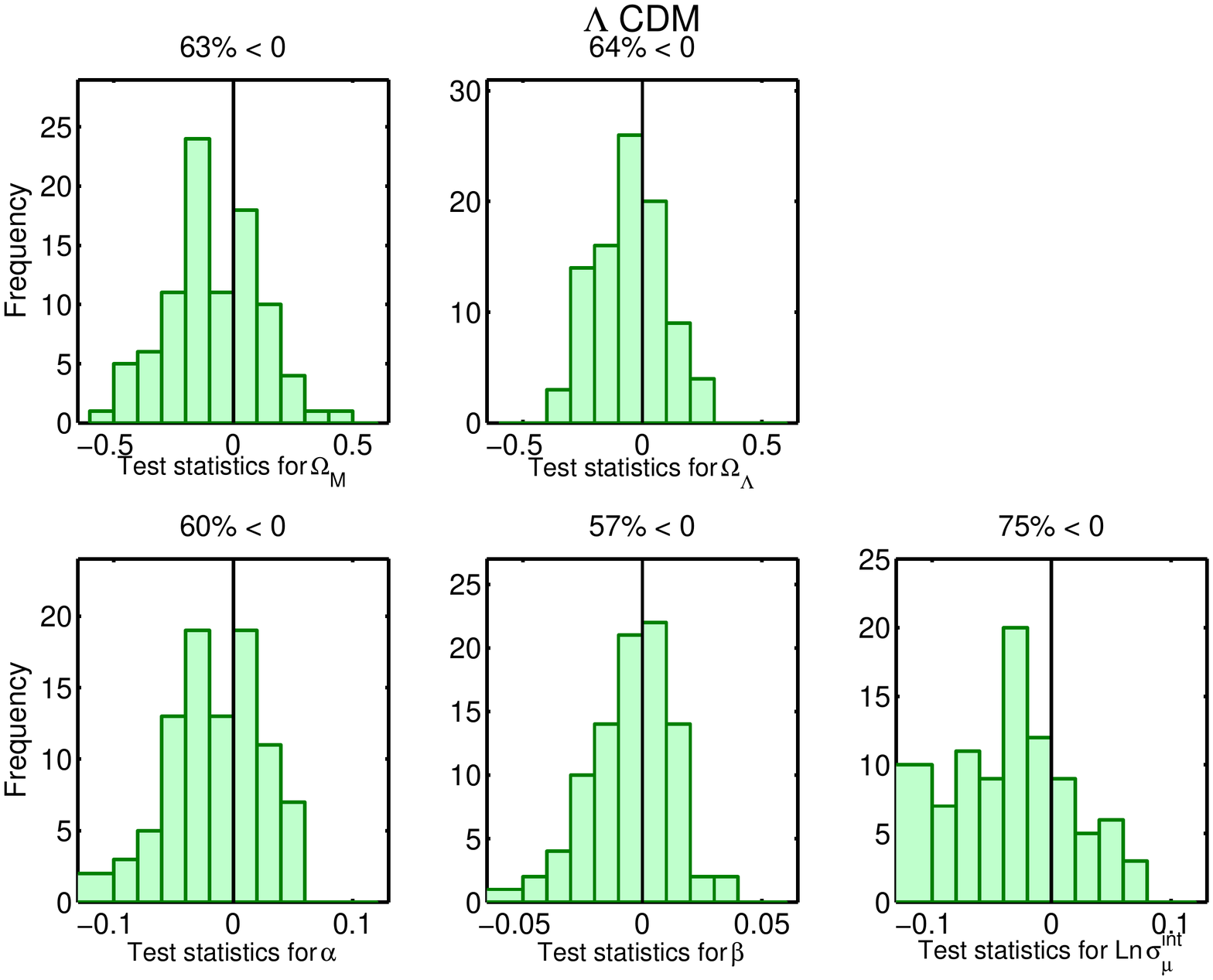} \hfill
\includegraphics[width=0.49 \linewidth]{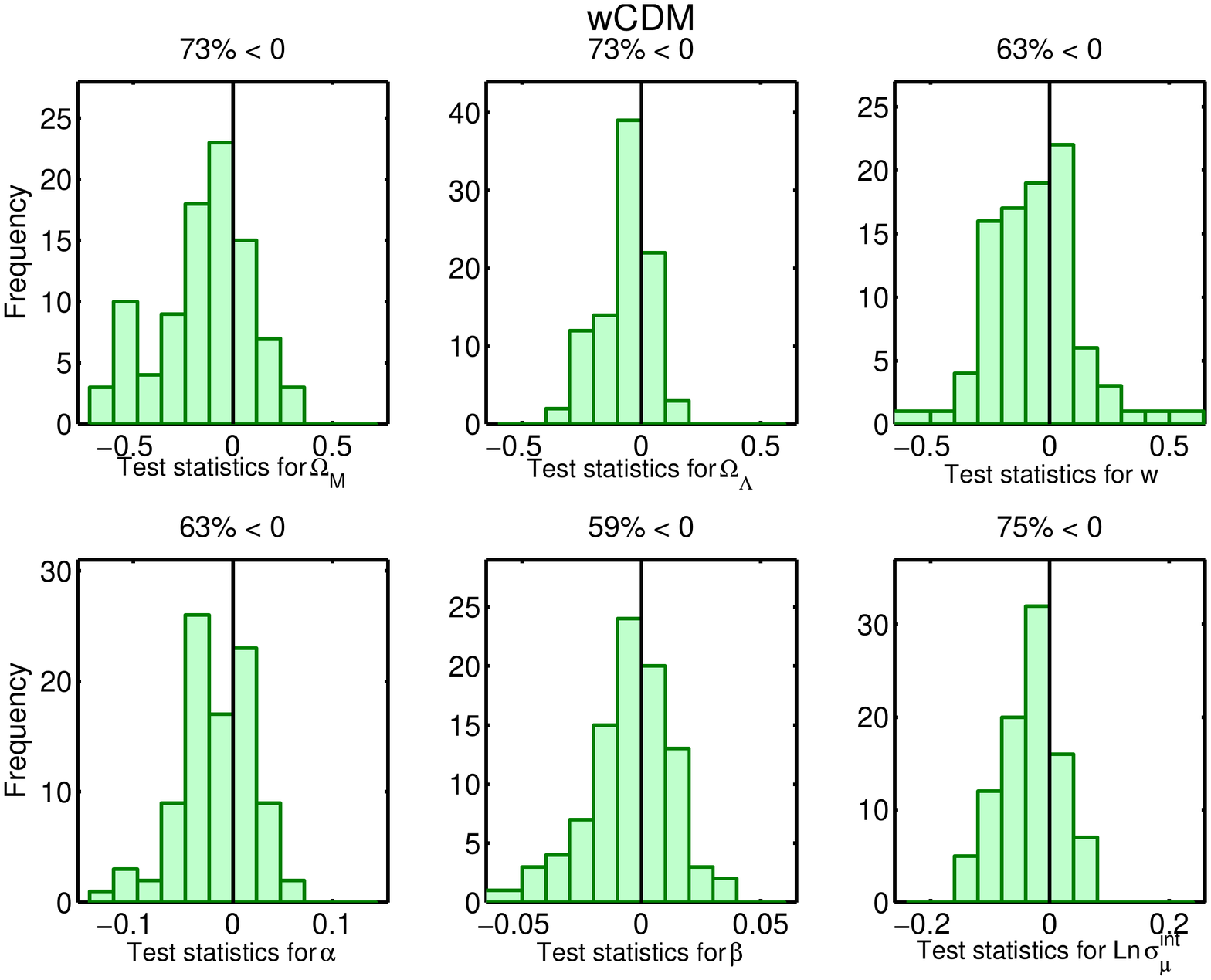} 
\caption{Histograms of the test statistics defined in Eq.~\eqref{eq:Dtheta}, comparing the long-term performance of the two methods for the parameters of interest in the $\Lambda$CDM model (left) and the wCDM model (right). A predominantly negative value of the test statistics means that our method gives a parameter reconstruction that is closer to the true value than the usual $\chisq$, i.e., less biassed. For the cosmological parameters (top row), our method outperforms $\chi^2$ about 2 times out of 3.}
\label{fig:mix7}
\end{figure*}

Tight errorbars are good, but not if they come at the expense of a biased reconstruction. To investigate this aspect, we build the following test statistics from each reconstruction: 
\be \label{eq:Dtheta}
\diff_\theta \equiv | \overline{\theta}_\text{BHM}/\theta_\text{true} -1| - |\theta^\text{bf}_{\chi^2}/\theta_\text{true}-1|, 
\ee
where $\overline{\theta}_\text{BHM}$ is the posterior mean recovered using our BHM method, $\theta^\text{bf}_{\chi^2}$ is the best-fit value for the parameter recovered using the standard $\chisq$ approach and $ \theta_\text{true}$ is the true value for that parameter. The meaning of $\diff_\theta$ is the following: for a given data realization, if the reconstructed posterior mean from our BHM is closer to the true parameter value than the best-fit $\chi^2$, then $\diff_\theta < 0$, which means that our method is less biassed than $\chi^2$. A histogram of the distribution of $\diff_\theta$ across the 100 realizations, shown in  Fig.~\ref{fig:mix7}, can be used to compare the two methods:  a negative average in the histogram means that the BHM outperforms the usual $\chisq$.  For all of the parameters considered, our method is clearly much less biassed than the $\chi^2$ method, outperforming $\chisq$ about 2/3 of the time. Furthermore, the reconstruction of the intrinsic dispersion is better with our Bayesian method almost 3 times out of 4. We emphasize once more that our methodology also provides an estimate of the uncertainty in the intrinsic dispersion, not just a best-fit value as the $\chi^2$ approach. 

We can further quantify the improvement in the statistical reconstruction by looking at the bias and mean squared error (MSE) for each parameter, defined as 
\begin{align}
\text{bias} & = \langle  \hat{\theta} - \theta_\text{true} \rangle \\ 
\text{MSE} &= \text{bias}^2 + \text{Var}, 
\end{align}
respectively, where the expectation is taken by averaging over the observed values in our 100 simulated trials, $\hat{\theta} = \overline{\theta}_\text{BHM}$ ($\hat{\theta} = \theta^\text{bf}_{\chi^2})$ for the BHM (for the $\chisq$ approach) and $\text{Var}$ is the observed parameter variance. The bias is the expectation value of the difference between estimator and true value, while the MSE measures the average of the squares of the errors, i.e., the amount by which the estimator differs from the true value for each parameter. Obviously, a smaller bias and a smaller MSE imply a better performance of the method. The results for the two methods are summarized in Table~\ref{tab:bmse}, which shows how our method reduces the bias by a factor $\sim 2-3$ for most parameters, while reducing the MSE by a factor of $\sim 2$. The only notable exception is the bias of the EOS parameter $w$, which is larger in our method than in the $\chisq$ approach. %This can be traced back to boundary effect of the prior, which cuts the Bayesian posterior at $w > -2$. 

\begin{figure*}
\centering
\includegraphics[width=0.49 \linewidth]{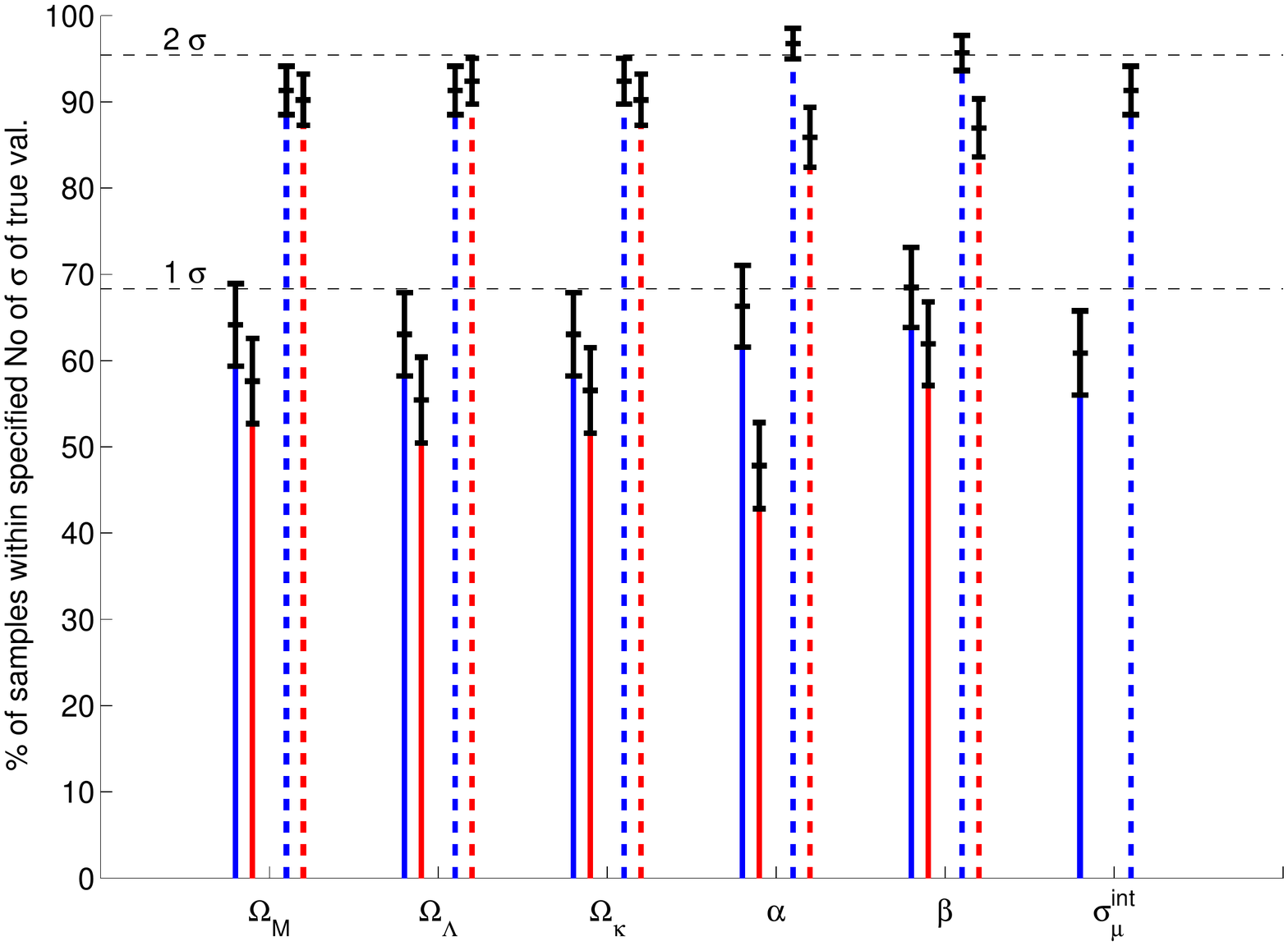} \hfill
\includegraphics[width=0.49 \linewidth]{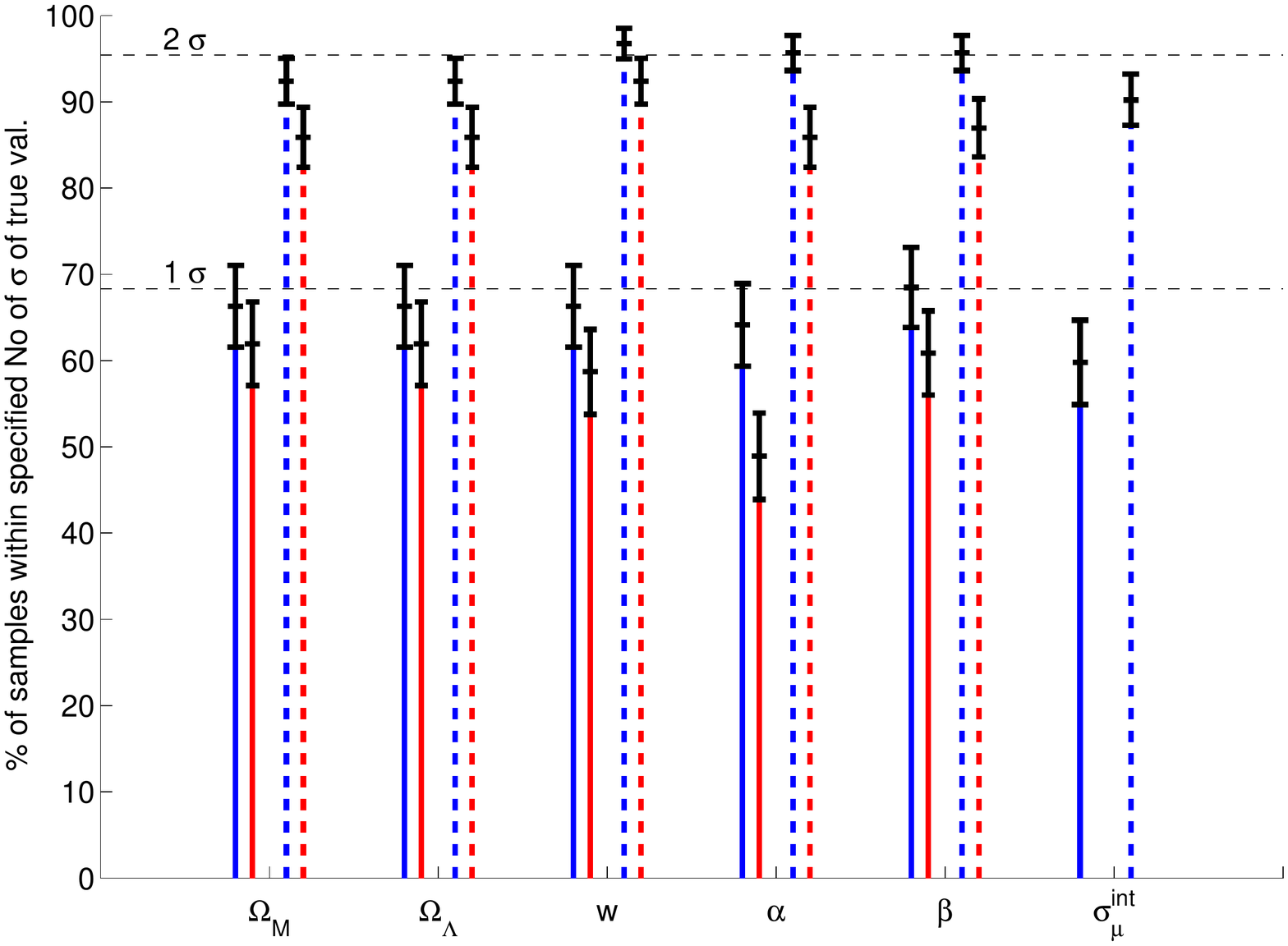} 
\caption{Coverage of our method (blue) and standard $\chi^2$ (red) for 68\% (solid) and 95\% (dashed) intervals, from 100 realizations of pseudo-data for the $\Lambda$CDM model (left) and the wCDM model (right). While both methods show significant undercoverage for all parameters, our method has a comparable coverage to the standard $\chi^2$, except for $w$. Coverage values for the intrinsic dispersion $\sigint$ are not available from the $\chi^2$ method, as it does not produce an error estimate for this quantity.
}
\label{fig:mix6}
\end{figure*}

Finally, in Fig.~\ref{fig:mix6} we plot the coverage of each method for 68\% and 95\% intervals. Errobars give an estimate of the uncertainty of the coverage result, by giving the binomial sampling error from the finite number of realizations considered, evaluated from the binomial variance as $Np(1-p)$, where $N=100$ is the number of trials and $p$ is the observed fractional coverage. Both method slightly undercover, i.e.~the credible region and confidence intervals are too short, although the lack of coverage is not dramatic: e.g., the typical coverage of the 1$\sigma$ (2$\sigma$) intervals from our method is $\sim 60\%$ ($90\%$). Our method shows slightly better coverage properties than the $\chi^2$ method, while producing considerably tighter and less biassed constraints (as demonstrated above). This further proves that the tighter intervals recovered by our method do not suffer from bias w.r.t the true values. 
%The only notable exception is the coverage for the dark energy equation of state $w$, which at $1\sigma$ in our method is only about 30\% (half of the coverage from the $\chi^2$ method). However, the $2\sigma$ coverage for $w$ is comparable from both methods. 

%---x---X---BIAS AND MSE TABLE ---X---x---
\begin{table}
\centering
\begin{tabular}{llllllll}
\hline\hline 
      & Parameter    &\multicolumn{3}{l}  {Bias}   & \multicolumn{3}{l}  {Mean squared error}   \\ 
      &              & Bayesian    &$\chi^2$& Improvement &  Bayesian   &$\chi^2$  & Improvement\\ \hline
$\Lambda$CDM & $\Om$ & -0.0188& -0.0183&1.0 &0.0082& 0.0147& 1.8 \\
      & $\OL$        & -0.0328& -0.0223& 0.7  &0.0307& 0.0458& 1.5 \\
      & $\alpha$     &  0.0012&  0.0032& 2.6 &0.0001& 0.0002& 1.4 \\
      & $\beta$      &  0.0202&  0.0482&2.4   &0.0118& 0.0163& 1.4\\
      & $\sigint$    & -0.0515& -0.1636&3.1  &0.0261& 0.0678& 2.6\\
\hline
$w$CDM& $\Om$        &-0.0177&-0.0494& 2.8 & 0.0072 &0.0207 & 2.9 \\
      & $\OL$        & 0.0177& 0.0494& 2.8 & 0.0072 &0.0207 & 2.9 \\
      & $w$          &-0.0852&-0.0111& 0.1 & 0.0884 &0.1420 & 1.6\\
      & $\alpha$     & 0.0013& 0.0032& 2.5 & 0.0001 &0.0002 & 1.5\\
      & $\beta$      & 0.0198& 0.0464& 2.3 & 0.0118 &0.0161 & 1.4\\
      & $\sigint$    &-0.0514&-0.1632& 3.2 & 0.0262 &0.0676 & 2.6 \\
\hline \hline
\end{tabular}
\caption{Comparison of the bias and mean squared error for our Bayesian method and the usual $\chisq$ approach. The columns labelled ``Improvement'' give the factor by which our Bayesian method reduces the bias and the MSE w.r.t.~the $\chisq$ approach. }
\label{tab:bmse}
\end{table}
%---x---X---BIAS AND MSE TABLE ---X---x---

%-----------------------FIGURE BLUE Hist b-----------------------------
%-----------------------FIGURE BLUE Hist b-----------------------------

%-----------------------FIGURE Green Hist -----------------------------
%-----------------------FIGURE Green Hist -----------------------------

%-----------------------FIGURE Coverage -----------------------------
%-----------------------FIGURE Coverage -----------------------------
%

%
%

\section{Cosmological constraints from current SNIa data}
\label{sec:current}

We now apply our BHM to fitting real SN data. We use the \salt{} fits result for 288 SNIa from~\cite{Kessler2009Firstyear}, which have been derived from 5 different surveys. Our method only includes statistical errors according to the procedure described in section~\ref{sec:bhm}, coming from redshift uncertainties (arising from spectroscopic errors and peculiar velocities), intrinsic dispersion (which is determined from the data) and full error propagation of the \salt{} fit results. Systematic uncertainties play an important role in SNIa cosmology fitting, and (although not included in this study) can also be treated in our formalism in a fully consistent way. We comment on this aspect further below, though we leave a complete exploration of systematics with our BHM to a future, dedicated work~\citep{March2}. 

\begin{figure*}
\centering
\includegraphics[width=0.48\linewidth]{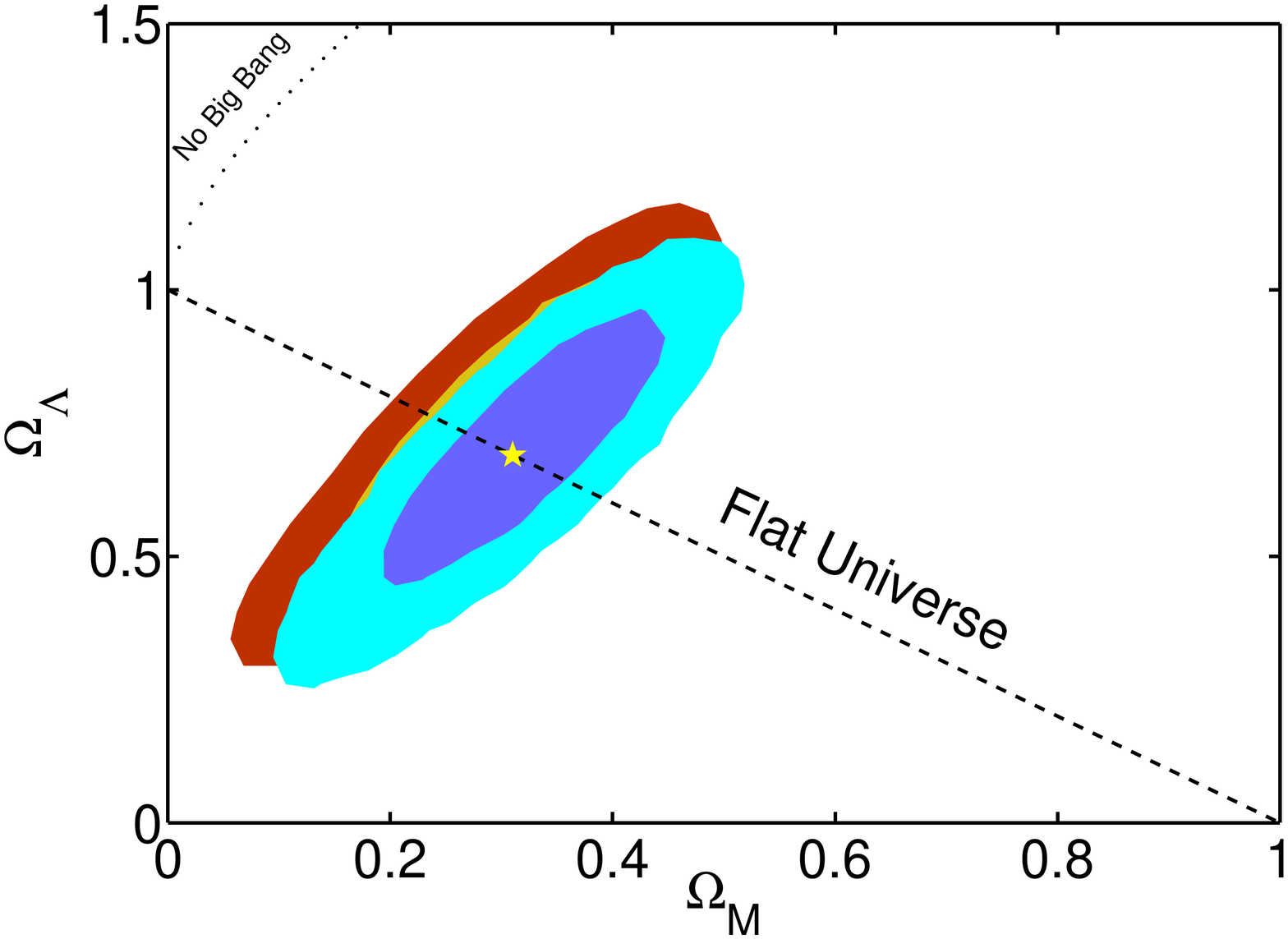}\hfill
\includegraphics[width=0.48\linewidth]{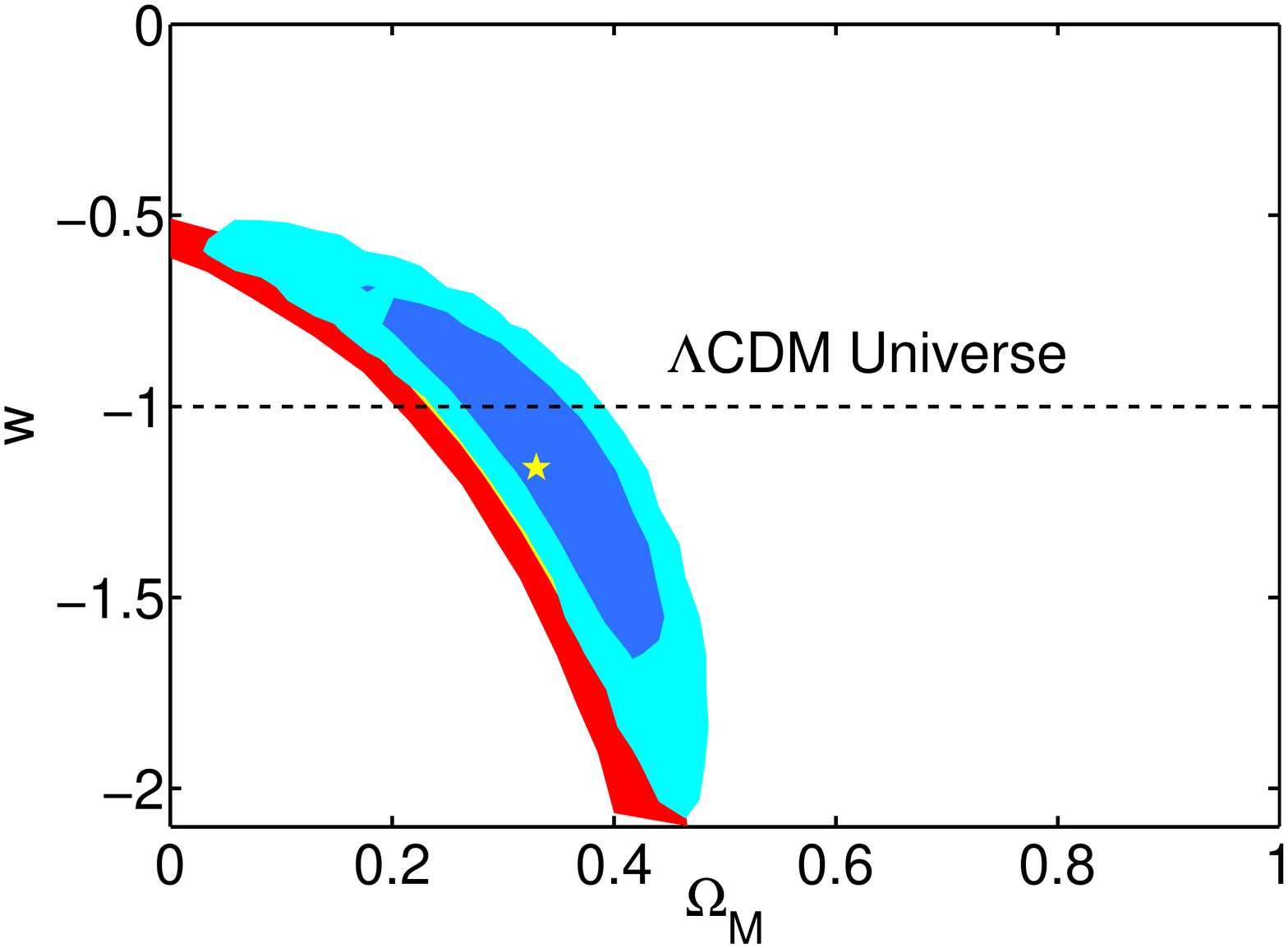}
\caption{Constraints on the cosmological parameters $\Om, \OL$ (left panel, assuming $w=-1$) and $w,\Om$ (right panel, assuming $\Omk = 0$) from our Bayesian method (light/dark blue regions, 68\% and 95\% marginalized posterior), compared with the statistical errors from the usual $\chi^2$ approach (yellow/red regions, same significance level; from Kessler at al. (2009a)). The yellow star gives the posterior mean from our analysis. 
\label{fig:real_data_combined_2D}
\label{fig:real_data_2D}}
\end{figure*}

We show in Fig.~\ref{fig:real_data_2D} the constraints on the cosmological parameters $\Om-\OL$ (left panel, assuming $w=-1$) and $w-\Om$ (right panel, assuming $\Omk = 0$) obtained with our method. All other parameters have been marginalized over. In order to be consistent with the literature, we have taken a non-informative prior on $H_0$, uniform in the range $[20,100]$ km/s/Mpc. The figure also compares our results with the statistical contours from~\cite{Kessler2009Firstyear}, obtained using the $\chi^2$ method. (Notice that we compare with the contours including only statistical uncertainties for consistency.)  In Fig.~\ref{fig:real_data_combined_2D} we combine our SNIa constraints with Cosmic Microwave Background (CMB) data from WMAP 5-yrs measurements~\citep{Komatsu:2008hk} and Baryonic Acoustic Oscillations (BAO) constraints from the Sloan Digital Sky Survey LRG sample~\citep{Eisenstein:2005su}, using the same method as \cite{Kessler2009Firstyear}. The combined SNIa, CMB and BAO statistical constraints result in $\Om=0.28 \pm 0.02, \OL=0.73 \pm 0.01$ (for the \lcdm{} model) and $\Om=0.28 \pm 0.01, w=-0.90 \pm 0.05$ (68.3\% credible intervals) for the wCDM model. Although the statistical uncertainties are comparable to the results by~\cite{Kessler2009Firstyear} from the same sample, our posterior mean values present shifts of up to $\sim 2\sigma$ compared to the results obtained using the standard $\chisq$ approach. This is a fairly significant shift, which can be attributed to our improved statistical method, which exhibits a reduced bias w.r.t. the $\chisq$ approach.

Fig.~\ref{fig:real_data_1D} shows the 1d marginalized posterior distributions for the Phillips correction parameters and for the intrinsic dispersion. All parameters are well constrained by the posterior, and we find $\alpha = 0.12 \pm 0.02$, $\beta = 2.7 \pm 0.1$ and a value of the intrinsic dispersion (for the whole sample) $\sigint = 0.13 \pm 0.01$ mag.

\cite{Kessler2009Firstyear} find values for the intrinsic dispersion ranging from 0.08 (for SDSS-II) to 0.23 (for the HST sample), but their $\chi^2$ method does not allow them to derive an error on those determinations. With our method, it would be easy to derive constraints on the intrinsic dispersion of each survey -- all one needs to do is to replace Eq.~\eqref{eq:Mi_normal} with a corresponding expression for each survey. This introduces one pair of population parameters $(M_0, \sigint)$ for each survey. In the same way, one could study whether the intrinsic dispersion evolves with redshift. We leave a detailed study of these issues to a future work. 

The value of $\alpha$ found in \cite{Kessler2009Firstyear}  is in the range $0.10-0.12$, depending of the details of the assumptions made, with a typical statistical uncertainty of order $\sim 0.015$. These results are comparable with our own. As for the colour correction parameter $\beta$, constraints from \cite{Kessler2009Firstyear} vary in the range $2.46-2.66$, with a statistical uncertainty of order $0.1-0.2$. This stronger dependence on the details of the analysis seems to point to a larger impact of systematic uncertainties for $\beta$, which is confirmed by evidence of evolution with redshift of the value of $\beta$ (\cite{Kessler2009Firstyear}, Fig.~39). Our method can be employed to carry out a rigorous assessment of the evolution with redshift of colour corrections. A possible strategy would be to replace $\beta$ with a vector of parameters $\beta_1, \beta_2, \dots$, with each element describing the colour correction in a different redshift bin. The analysis proceeds then as above, and it produces posterior distributions for the components of $\beta$, which allows to check the hypothesis of evolution. Finally, in such an analysis the marginalized constraints on all other parameters (including the cosmological parameters of interest) would automatically include the full uncertainty propagation from the colour correction evolution, without the need for further {\em ad hoc} inflation of the errorbars. These kind of tests will be pursued in a forthcoming publication~\citep{March2}.

\begin{figure*}
\centering
\includegraphics[width=0.48\linewidth]{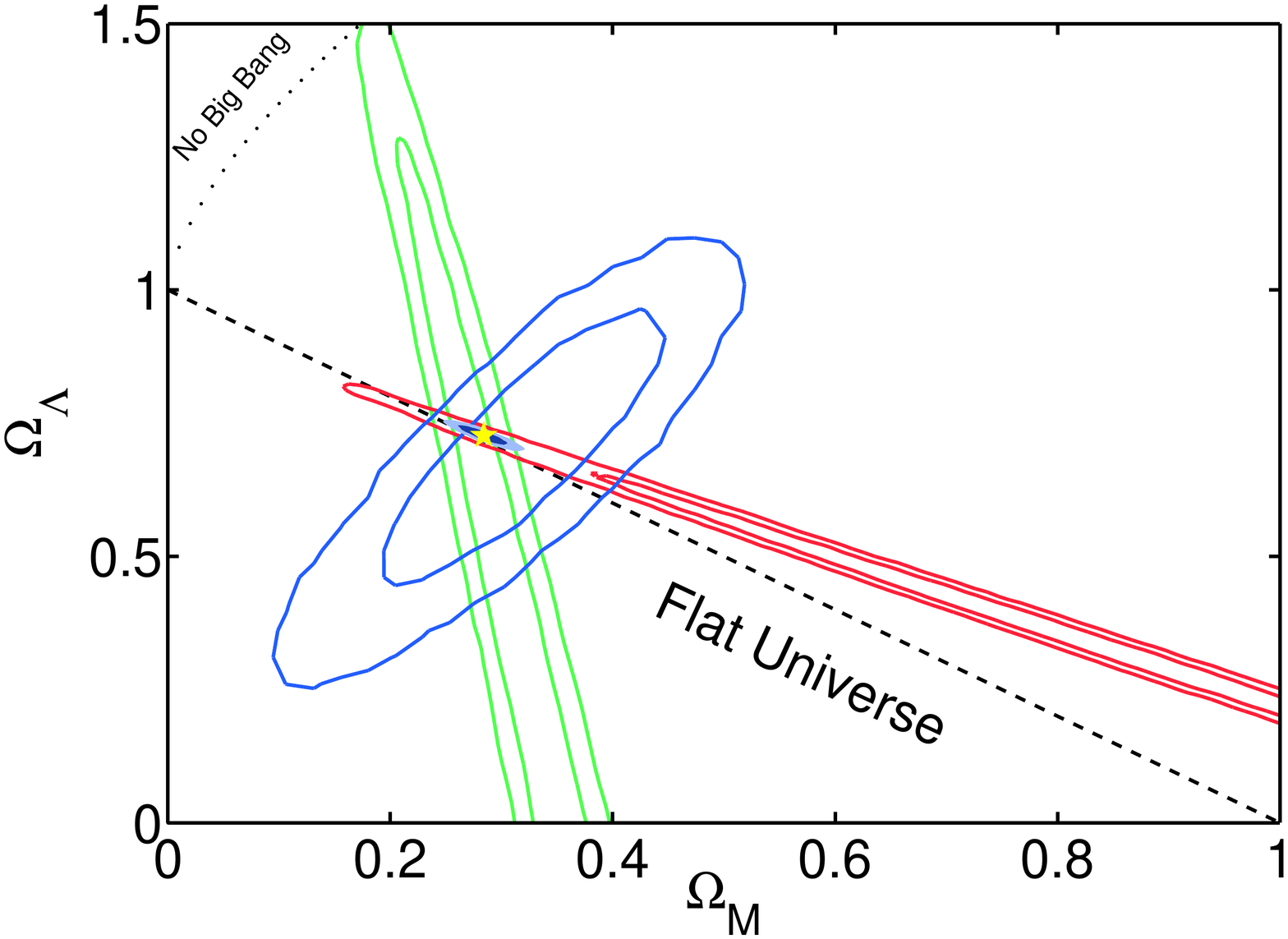}\hfill
\includegraphics[width=0.48\linewidth]{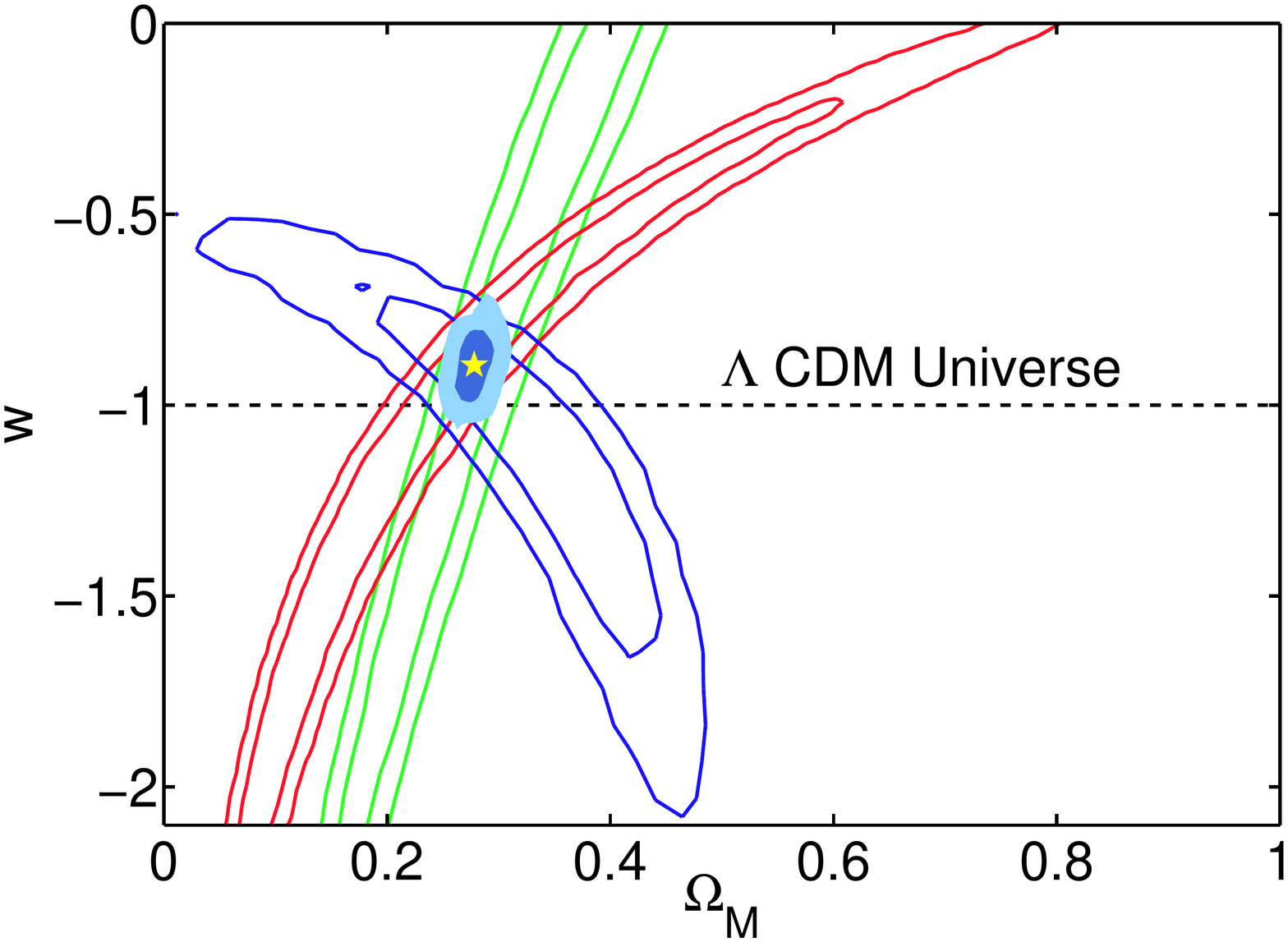}
\caption{Combined constraints on the cosmological parameters $\Om,\OL$ (left panel, assuming $w=-1$) and $w,\Om$ (right panel, assuming $\Omk = 0$) from SNIa, CMB and BAO data. Red contours give 68\% and 95\% regions from CMB alone, green contours from BAO alone, blue contours from SNIa alone from our Bayesian method. The filled regions delimit  68\% and 95\% combined constraints, with the yellow star indicating the posterior mean. \label{fig:real_data_combined_2D}}
\end{figure*}

\begin{figure*}
\centering
\hfill
\includegraphics[width=0.33\linewidth]{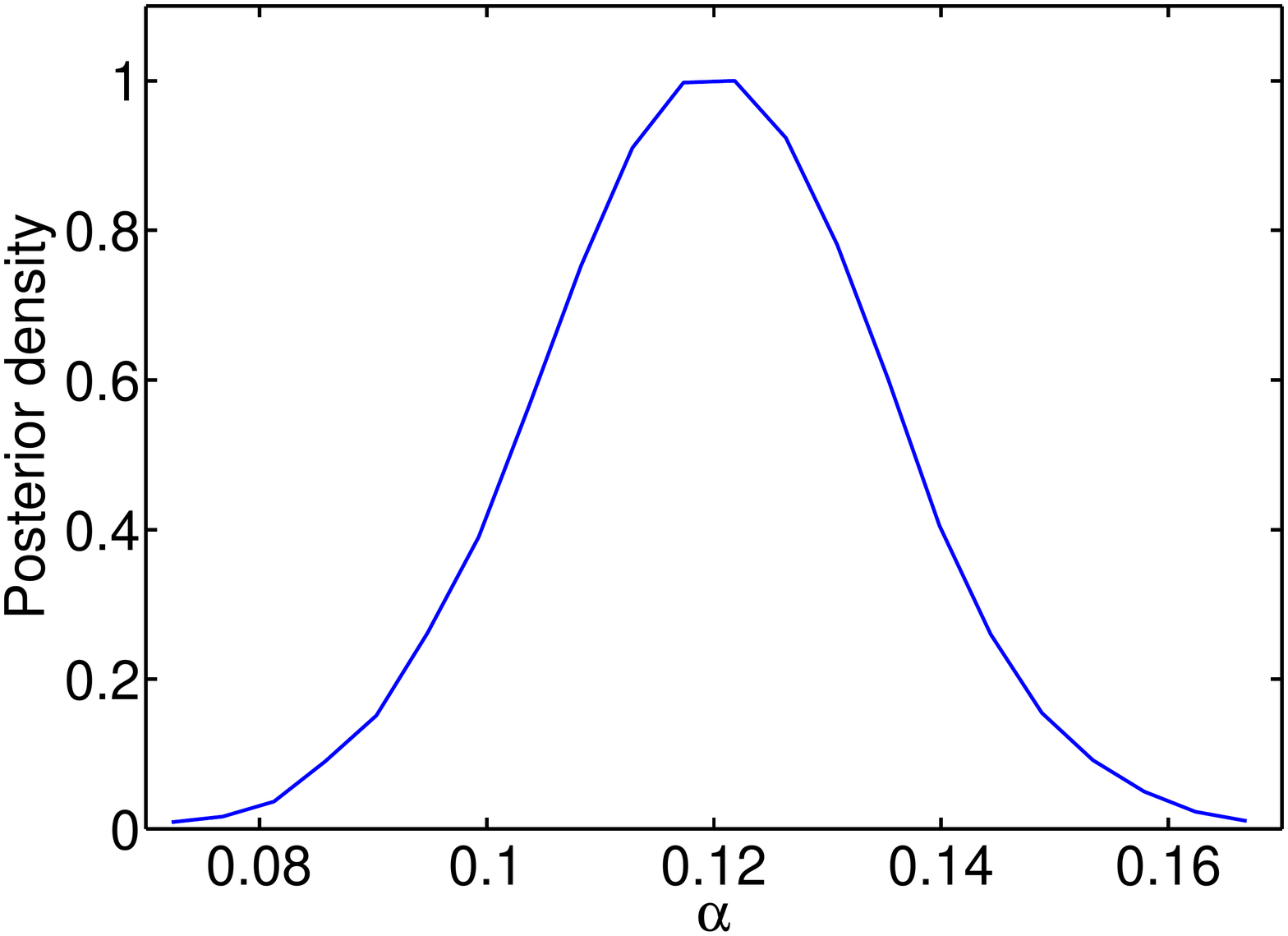} \hfill
\includegraphics[width=0.33\linewidth]{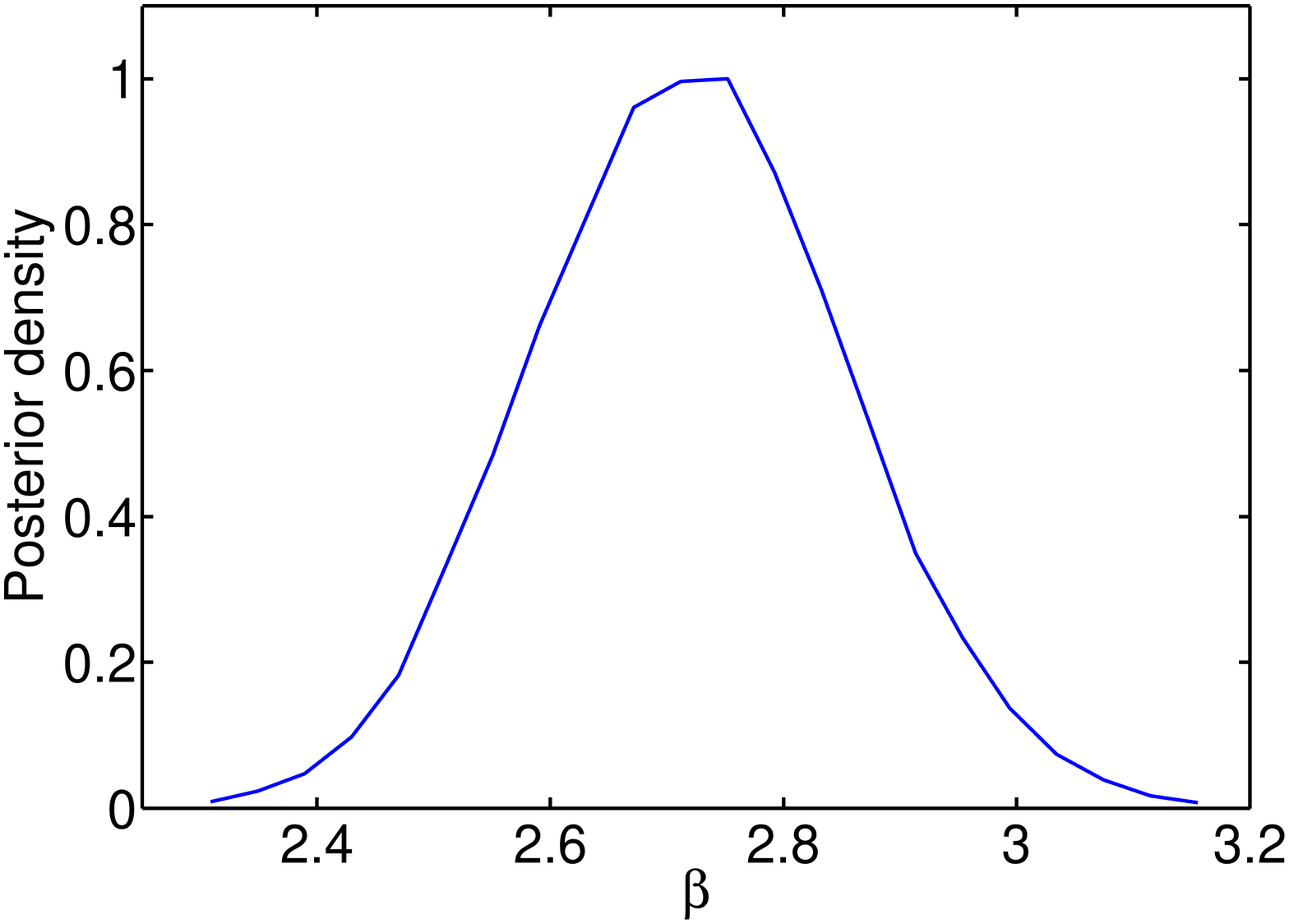} \hfill
\includegraphics[width=0.33\linewidth]{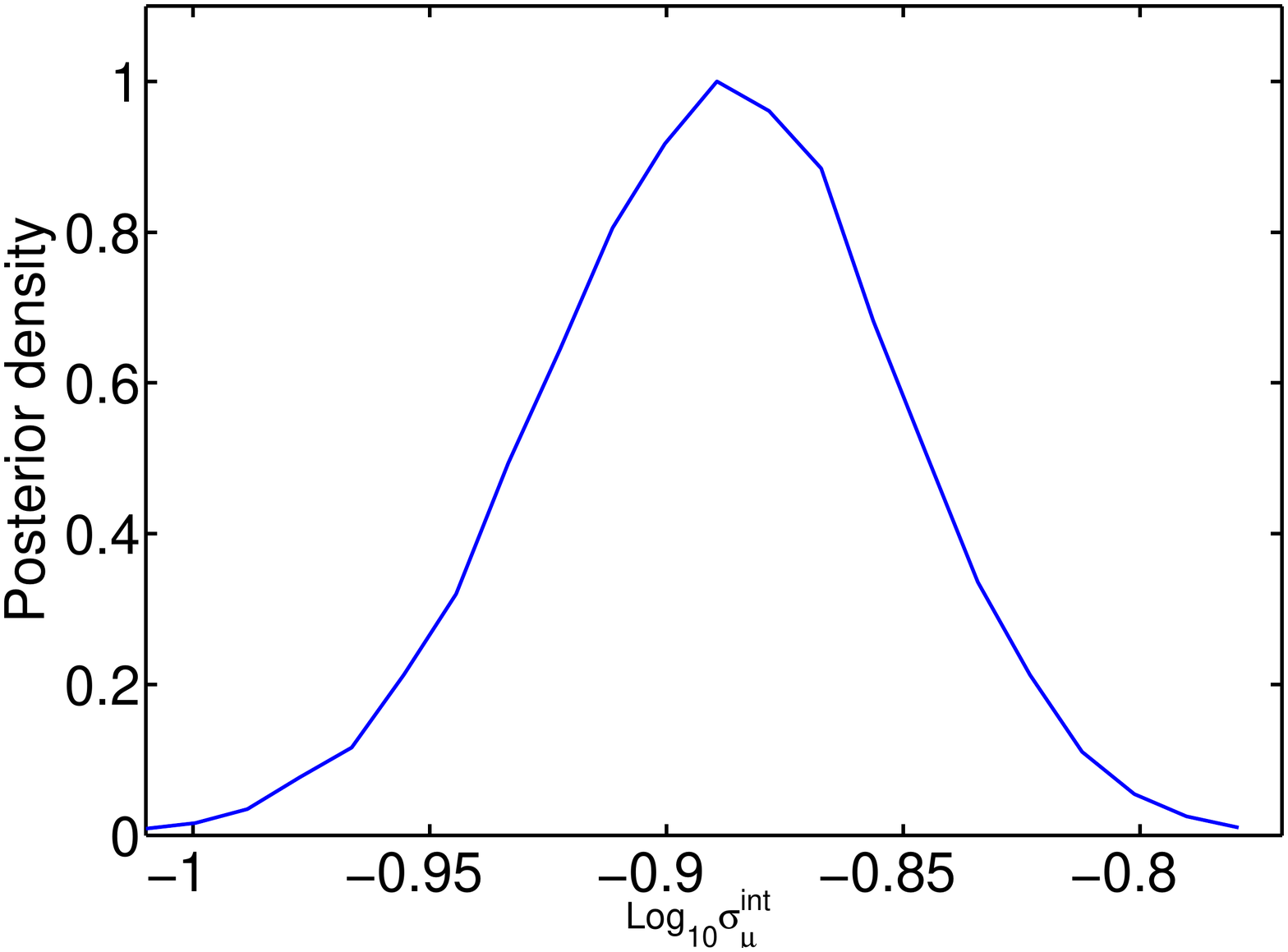} 
\caption{Marginalised posterior for the stretch correction $\alpha$, colour correction $\beta$ parameter and logarithm of the intrinsic dispersion of SNe, $\log\sigint$ from current SNIa data.}
\label{fig:real_data_1D}
\end{figure*}

\section{Conclusions}
\label{sec:conclusions}

We have presented a statistically principled approach for the rigorous analysis of \salt{} SNIa lightcurve fits, based on a Bayesian hierarchical model. The main novelty of our method is that it produces an effective likelihood that propagates uncertainties in a fully consistent way. We have introduced an explicit statistical modeling of the intrinsic magnitude distribution of the SNIa population, which for the first time allows one to derive a full posterior distribution of the SNIa intrinsic dispersion. 

We have tested our method using simulated data sets and found that it compares favourably with the standard $\chi^2$ approach, both on individual data realizations and in the long term performance. Statistical constraints on cosmological parameters are significantly improved, while in a series of 100 simulated data sets our method outperforms the $\chi^2$ approach at least 2 times out of 3 for the parameters of interest. We have also demonstrated that our method is less biassed and has better coverage properties than the usual approach.

We applied our methodology to a sample of 288 SNIa from multiple surveys. We find that the flat $\Lambda$CDM model is still in good agreement with the data, even under our improved analysis. However, the posterior mean for the cosmological parameters exhibit up to $2\sigma$ shifts w.r.t. results obtained with the conventional $\chisq$ approach. This is a consequence of our improved statistical analysis, which benefits from a reduced bias in estimating the parameters.

While in this paper we have only discussed statistical constraints, our method offers a new, fully consistent way of including systematic uncertainties in the fit. As our method is fully Bayesian, it can be used in conjunction with fast and efficient Bayesian sampling algorithms, such as MCMC and nested sampling. This will allow to enlarge the number of parameters controlling systematic effects that can be included in the analysis, thus taking SNIa cosmological parameter fitting to a new level of statistical sophistication. The power of our method as applied to systematic errors analysis will be presented in a forthcoming, dedicated work.
 
At a time when SNIa constraints are entering a new level of precision, and with a tenfold increase in the sample size expected over the next few years, we believe it is timely to upgrade the cosmological data analysis pipeline in order to extract the most information from present and upcoming SNIa data. This paper represents a first step in this direction.  

\bigskip
\textit{Acknowledgments} This work was partially supported by travel grants by the Royal
Astronomical Society and by the Royal Society. RT would like to thank the Volen Center for Complex Systems and the African Institute for Mathematical Sciences for hospitality. MCM would like to thank CWRU for hospitality. PMV, GDS and PB would like to thank Imperial College for hospitality. MCM was partially supported by a Royal Astronomical Society grant. GDS and PV were supported by a grant from the US-DOE to the CWRU theory
group, and by NASA grant NNX07AG89G to GDS.  PV was supported by CWRU's
College of Arts and Sciences.
We would like to thank Bruce Bassett, Ariel Goobar, Josh Frieman, Andrew Jaffe, Marek Kowalski, Tom Loredo, Louis Lyons, Kaisey Mandel, Mathew Smith, Mark Sullivan, for useful discussions and Alex Conley, Rick Kessler, David van Dyke for comments on an earlier draft. The use of Imperial College London's High Performance Computing service is gratefully acknowledged.

%People we like to cite: \\

%KesslerBecker2009 \cite{KesslerBecker2009} \\
%NordinGoobar2008 \cite{NordinGoobar2008} \\

%\bibliographystyle{mn2e}
%\bibliography{BasicBiblio,AdditionalRefs,rxt-sn}

\newcommand{\bibfont}{\fontsize{10}{12}\selectfont} \newcommand{\noopsort}[1]{}
  \newcommand{\printfirst}[2]{#1} \newcommand{\singleletter}[1]{#1}
  \newcommand{\switchargs}[2]{#2#1}

\appendix

\section{Notation} \label{sec:notation}

For reference, we collect here a few useful formulas relating to Gaussian integrals. Use the notation $\ul{x} \sim \norm_{\ul{x}}(\ul{\mu}, \Sigma)$ to denote that the random variable $\ul{x}$ is drawn from a Normal distribution of mean $\ul{\mu}$ and inverse covariance matrix $\Sigma$, given by
\be \label{eq:gaussmd}
p(\ul{x}) = \frac{1}{|2\pi \Sigma|^{1/2}} \exp \left[-\frac{1}{2}(\ul{x} - \ul{\mu})^T \Sigma^{-1} (\ul{x} - \ul{\mu})\right].
\ee
In performing Gaussian integrals, it is also convenient to recall that:
\be \label{eq:Gaussian_combination}
\norm_{\ul{x}}(\ul{\mu}_1, \Sigma_1)\cdot \norm_{\ul{x}}(\ul{\mu}_2, \Sigma_2) = f_0 \cdot \norm_{\ul{x}}(\ul{\mu}_f, \Sigma_f)
\ee
where 
\begin{align}
f_0 & = \norm_{\ul{\mu}_1}(\ul{\mu}_2, (\Sigma_1+\Sigma_2)) \\
\ul{\mu}_f & = (\Sigma_1^{-1} + \Sigma_2^{-1})^{-1}(\Sigma_1^{-1}\ul{\mu}_1 + \Sigma_2^{-1}\ul{\mu}_2) \\
 \Sigma_f & =  (\Sigma_1^{-1} + \Sigma_2^{-1})^{-1} .
\end{align}
Finally, the integral of the multidimensional Gaussian of Eq.~\eqref{eq:gaussmd} over all space is unity.

\section{Toy model for general linear fitting}
\label{app:toy}

We can get an intuitive understanding of the central ingredient in our Bayesian hierarchical method by considering a simpler toy model, which highlights the salient features of the problem. 

As mentioned in section~\ref{sec:chisq}, several authors have observed in the past that the spread of fit values for the colour correction, $\cbh$, is of the same order as the size of the statistical uncertainty on the values themselves. This leads to a bias in the best-fit value of $\beta$ obtained from minimizing the $\chi^2$ in Eq.~\eqref{eq:chisq}. The quantity $\beta$ gives the slope of the linear relationship between $\cbl$ and $\muu$, see Eq.~\eqref{eq:phil2vec}. In the usual $\chi^2$ approach, the latent (true) $\cbl$ are replaced by the observed value, $\cbh$, as in Eq.~\eqref{eq:mu_relaced}. If the statistical uncertainty in the independent variable, $\cbh$, is as large as the spread of its values, the best-fit slope obtained from minimizing the $\chi^2$ may be biased, as the large uncertainty in the actual location of the independent variable leads to confusion as to the value of the slope. 
The (Bayesian) solution to this problem is to determine the spread of the independent variable directly from the data, and to marginalize over it with an appropriate prior. This gives the most general solution to the problem of linear fitting with errors in both the independent and dependent variable, as shown by~\cite{Gull:1989}. Recent literature in cosmology and astronomy~\citep{DAgostini:2005pk,Hogg:2010yz} addresses linear fitting, but not this general case, which has been treated before in~\cite{Kelly:2007jy} (see also~\cite{Andreon:2006fe,Andreon:2010gu} for related examples). In this short appendix, we will analyse the toy model of fitting a linear function, and compare the performance of a BHM and the $\chi^2$ approach.

\subsection{Bayesian linear fitting in the presence of large $x$ and $y$ uncertainties}

\begin{figure}
\centering
\includegraphics*[angle = 0]{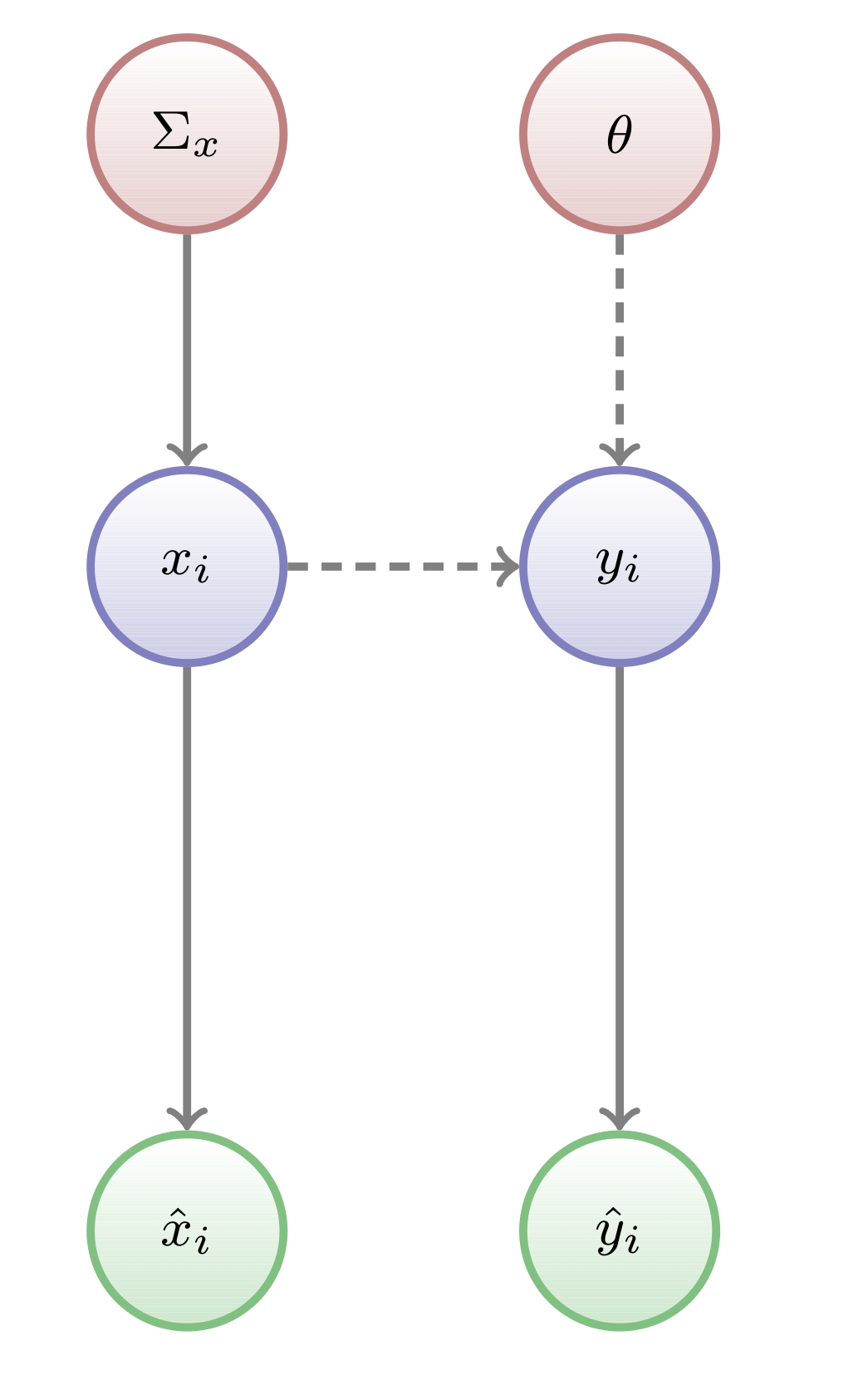}
\caption{Bayesian network showing the deterministic and probabilistic connections in the toy linear model. Solid lines indicate probabilistic connections, dashed lines represent deterministic connections. Parameters to be constrained are in red, latent variables in blue and data in green (denoted by hats). $\theta$ denotes the parameters ${a,b}$, i.e. the intercept and slope of the linear relation of Eq.~\eqref{eq:gull:linear}.}
\label{fig:gullbay}
\end{figure}

In this subsection, we give a short review of the results of \cite{Gull:1989}. The simplest toy model which illustrates the methodology we adopt in our paper is shown by the graphical network of Fig.~\ref{fig:gullbay}.  A linear relationship is assumed between the latent variables $x_i$ and $y_i$, described by a slope $a$ and intercept $b$ (which are collectively denoted as a parameter vector $\theta$): 
\be
\label{eq:gull:linear}
y_i = ax_i + b. 
\ee
The observed values for the dependent and independent variables are denoted by hats ($\hat{x}_i,\hat{y}_i$, $i=1,\dots,N$), and they are obtained from the latent values under the assumption of Gaussian noise (with known variances $\sx^2, \sy^2$): 
\be
\hat{x}_i \sim \norm(x_i, \sx^2) \, \text{ and } \hat{y}_i \sim \norm(y_i, \sy^2). 
\ee
This probabilistic relationship is depicted in Fig.~\ref{fig:gullbay} by the solid arrows connecting the latent variables to the observed quantities. Assuming that errors are uncorrelated, the joint likelihood is given by\footnote{For ease of notation, quantities without subscript $i$ denote in the following $N$-dimensional vectors, e.g. $x = \{ x_1, \dots, x_N \}$. }
\be \label{eq:gulllike}
p(\xh,\yh|x,y, \sx,\sy, \theta) = (4 \pi^2 \sx^2 \sy^2)^{-N/2} \exp \left(-\frac{1}{2} \[  \frac{\sum_i(\xh_i -x_i)^2}{\sxs} +  \frac{\sum_i(\yh_i -y_i)^2}{\sys} \] \right).
\ee
The problem can be made more symmetric by defining rescaled versions of the data: 
\be
\hat{X} = \frac{\xh - x_0}{R_x} \, \text{ and } \hat{Y} = \frac{\yh - y_0}{R_y},
\ee
where the variables $x_0, y_0$ describe the mean value of $\xh, \yh$, while $R_x, R_y$ describe their spread. These new variables are related to the old ones ($a,b$) by 
\be \label{eq:b}
b = y_0 - a x_0\, \text{ and } a = R_y/R_x.
\ee
Neglecting the normalization constant, the joint posterior for $x,y, x_0, y_0, R_x, R_y$ can be written as:
\be \label{eq:gullpost}
p(x, y, x_0,y_0, R_x, R_y| \xh, \yh, \sx, \sy)  \propto p(\xh,\yh|x,x_0,y_0,R_x,R_y) p(x|x_0,y_0,R_x,R_y)p(x_0,y_0,R_x,R_y),
\ee
where the first term on the r.h.s.~is the likelihood of Eq.~\eqref{eq:gulllike}. The key step is to recognize that the appropriate conditional distribution for the latent $x$ is
\be
\label{eq:priorx}
p(x|x_0,y_0,R_x,R_y)=(4 \pi^2 R_x^2)^{-1/2}\exp \left(- \frac{1}{2} \frac{\sum_i (x_i-x_0)^2}{R_x^2} \right).
\ee
This describe a prior over $x$ centered around the hyperparameter $x_0$ and with standard deviation $R_x$. Crucially, both $x_0$ and $R_x$ are unknown, and are explicitly determined from the data in the joint posterior, before being marginalized out at the end. Finally, for the prior $p(x_0,y_0,R_x,R_y)$ appearing in Eq.~\eqref{eq:gullpost} we adopt a uniform prior on $x_0, y_0$ (as those are location variables) and a prior uniform in $\log R_x, \log R_y$ (those being scale variables, as apparent from \eqref{eq:priorx}). 

From here, one can eliminate $y$ from Eq.~\eqref{eq:gullpost} using Eq.~\eqref{eq:gull:linear}, then trade $b$ for $R_y$ using Eq.~\eqref{eq:b}, to obtain:
\be
p(x,x_0,y_0 , a ,\log R_x, \log R_y | \xh, \yh, \sx, \sy) \propto (8 \pi^3 \sx^2 \sy^2 R_x^2)^{-N/2}  \exp \left(-\frac{1}{2} \[  \frac{\sum_i(\xh_i -x_i)^2}{\sxs} 
 +  \frac{\sum_i(\yh_i -ax_i -y_0 +ax_0)^2}{\sys}   
+   \frac{\sum_i(x_i -x_0)^2}{R_x^2}  \]\right).
\ee
From this expression, the latent $x$ can be marginalized out analytically, as well as the nuisance parameters $x_0, y_0$, by using appropriate completions of the square in the Gaussian. After some algebra, one finally obtains 
\be \label{eq:post_toy_final}
p(a, \log R | \xh, \yh, \sx, \sy ) \propto (a^2 \sxs R_x^2 + \sxs \sys + \sys R_x^2 )^{-\frac{N-1}{2}}
 \exp \left(-\frac{1}{2}\frac{V_{xx}(a^2R_x^2 + \sys) -2 V_{xy} a R_x^2 + V_{yy}(R_x^2+\sxs)}{a^2 \sxs R_x^2 + \sxs \sys + \sys R_x^2 }\right)   
\ee
where:
\be
\label{eq:sxx}
V_{xx}^2=  \sum_i (\xh_i -\bar{x})^2, \quad 
V_{xy}^2= \sum_i (\xh_i -\bar{x})(\yh_i -\bar{y}), \quad 
V_{yy}^2=\sum_i (\yh_i -\bar{y})^2, \quad R = (R_x R_y)^{1/2}.
\ee
In the above, $\bar{x} = \sum_i \xh_i/N$, and similarly for $\bar{y}$. The marginal posterior for the slope $a$ is obtained by numerical marginalization of $\log R$ from the above expression.

%%%%%%%%%%%%%%%%%%%%%%%%%%%%%%%%%%%%%%%%%%%%%%%%%%%%%%%%%%%%%%%%%%%%%%%%%%%%%%%%%%%%%%%%%%%%%%%%%%

\subsection{Comparison with the $\chi^2$ approach} %---------------------------------------------------------------

If instead of the statistically principled solution found above, one writes down a simple $\chi^2$ expression for the likelihood, including error propagation from the linear relationship \eqref{eq:gull:linear} one would obtain~\citep{DAgostini:2005pk}: 
\be
\label{eq:chisqtoy} 
L(a,b) \propto \exp \left(- \frac{1}{2} \sum_i \frac{(\yh_i -a \xh_i - b)^2}{\sys +a^2\sxs}  \right),
\ee
from where the intercept $b$ is eliminated by maximising over it (profiling). We now compare the reconstruction of the slope parameter $a$ from Eq.~\eqref{eq:chisqtoy} (with $b$ eliminated via profiling) with the result obtained using the Bayesian expression, \eqref{eq:post_toy_final}, marginalized numerically over $\log R$ (with a uniform prior on $\log R$) with the help of simulated data.  

\begin{figure}
\centering
\includegraphics*[angle = 0, totalheight=7cm]{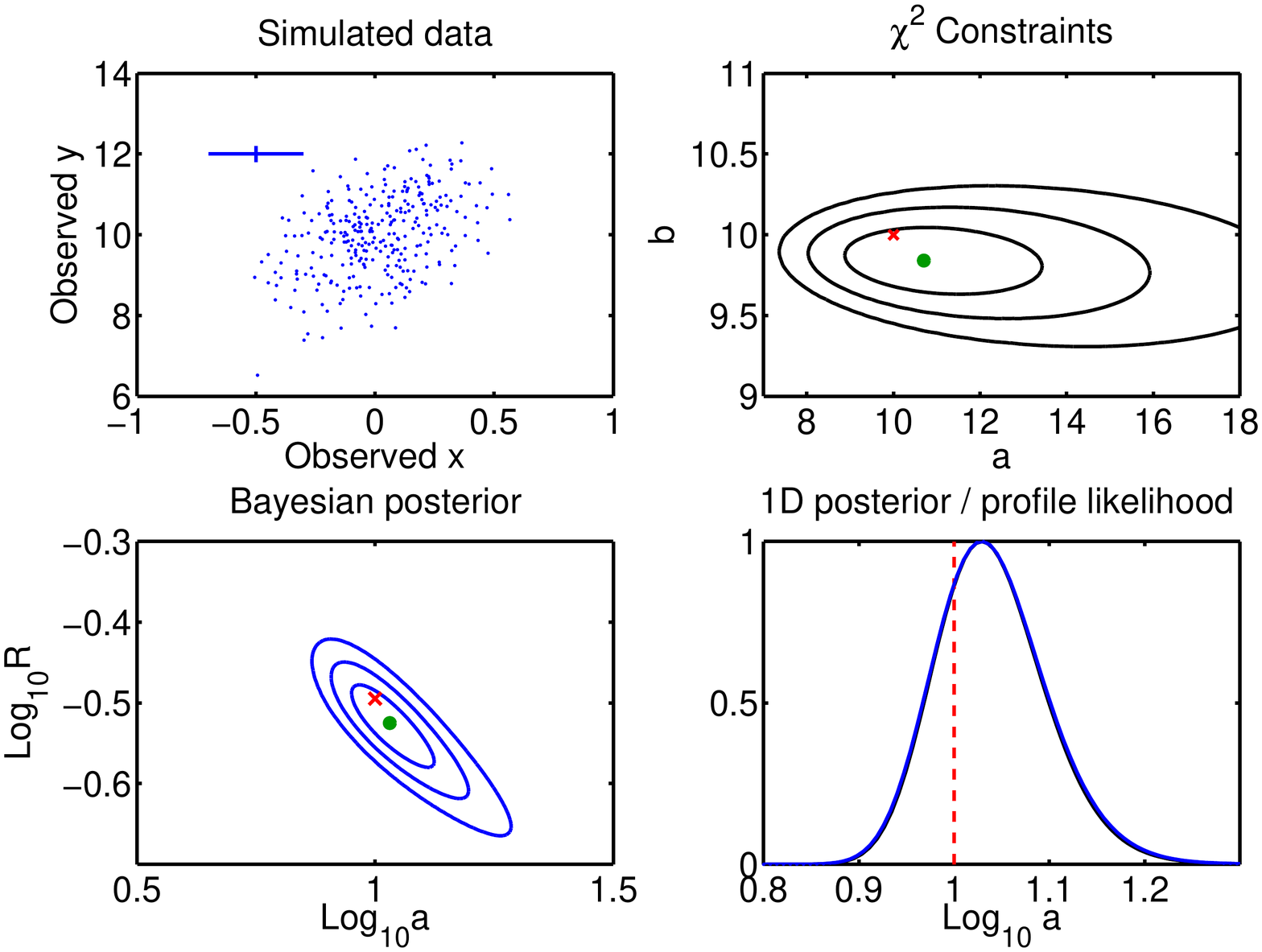}
\caption{Numerical comparison between the $\chi^2$ approach and the Bayesian method for linear fitting. Upper left panel: data set of $N=300$ observations with errors both in the $x$ and $y$ directions, given by the error bar. Upper right panel: reconstruction of the slope $a$ and intercept $b$ using a $\chi^2$ likelihood (red cross is the true value, green circle the maximum likelihood value). Lower left panel: Bayesian posterior (Eq.~\eqref{eq:post_toy_final}) in the $\log a$, $\log R$ plane, with green circle showing posterior mean. In both panels, contours enclose $1\sigma$, $2\sigma$ and $3\sigma$ regions. Lower right panel: marginalized Bayesian posterior (blue) and profiled $\chi^2$ likelihood (black, lying exactly on top of the blue curve), with the dashed line showing the true value. The two methods give essentially identical results in this case.}
\label{fig:toy1}
\end{figure}

\begin{figure}
\centering
\includegraphics*[angle = 0, totalheight=7cm]{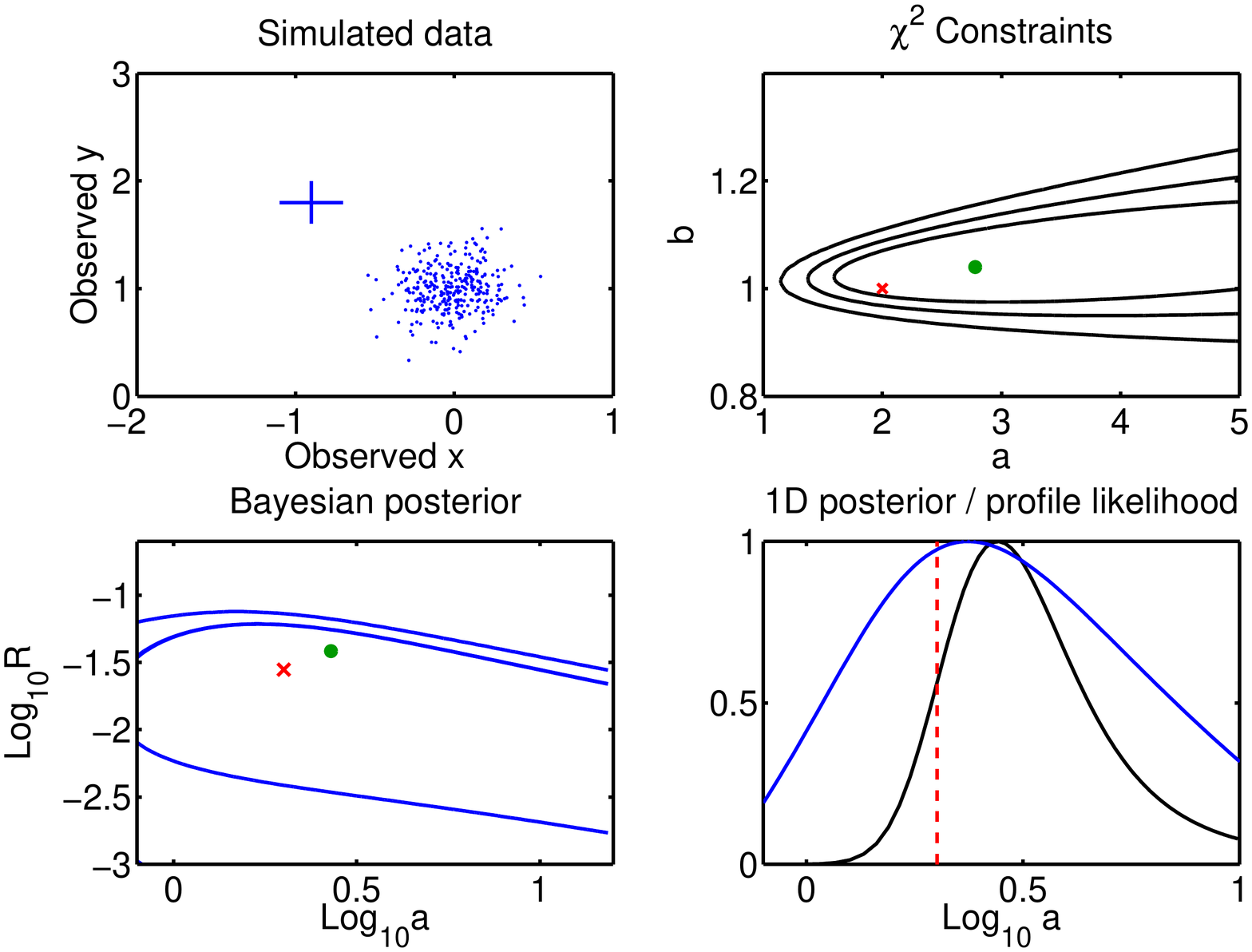}
\caption{As in Fig.~\ref{fig:toy2}, but now with a larger statistical uncertainty in the data compared to their spread. The Bayesian marginal posterior for the slope (blue, lower right panel) peaks much closer to the true value than the $\chi^2$ expression (black). In the lower left panel, the $3\sigma$ contour from the Bayesian method lies outside the range of the plot.}
\label{fig:toy2}
\end{figure}

Figs.~\ref{fig:toy1} and \ref{fig:toy2} show in the upper left panel the simulated data points ($N=300$), with the error bars giving the size of the standard deviation in the $x$ and $y$ directions for each datum (i.e., the value of $\sx, \sy$). The contour plots depict joint posterior regions for $(\log a, \log R)$ from the Bayesian expression of Eq.~\eqref{eq:post_toy_final}, and confidence regions from the $\chi^2$ likelihood~\eqref{eq:chisqtoy} in the $a,b$ plane. The lower right panel shows the 1d marginalized posterior distribution for $\log a$ (blue) and the profile likelihood from the $\chi^2$ approach (black). Fig.~\ref{fig:toy1} shows that the two results are essentially identical when the size of the statistical error is smaller than the spread of values of the data points (in this particular example, by a factor of 2). However, when the statistical uncertainty in the $x$ direction is as large as or larger than the spread of the points (Fig.~\ref{fig:toy2}), the $\chi^2$ approach gives a biassed result for the slope parameter $a$. The Bayesian posterior, on the contrary, is closer to the true value, although it (correctly) shows a larger uncertainty.

%\input{app/appendix_likelihood}
%!TEX root =  ../mtbsv.tex
\section{Derivation of the effective likelihood} \label{app:like}
\subsection{Integration over intrinsic redshifts}
In order to perform the multi-dimensional integral over $\zbl$, we Taylor expand $\muu$ around $\zbh$ (as justified by the fact that redshift errors are typically small: the error from 300 km/s peculiar velocity is $\sigma_{z_i} = 0.0012$, while the error from spectroscopic redshifts from SNe themselves is  $\sigma_{z_i} = 0.005$, see~\cite{Kessler2009Firstyear}):
\begin{equation} \label{eq:mu_taylor}
  \mu_j=\mu(z_j) = 5\log_{10} \left(\frac{D_{L}(z_j)}{\text{Mpc}}\right)+25 
  \approx \mu(\hat{z}_j) + 5(\log_{10}e) \left. \frac{\partial_{z_j} D_L(z_j)}{D_L(z_j)}\right|_{\hat{z}_j} (z_j - \hat{z}_j).
\end{equation}
With this approximation we can now carry out the multi-dimensional integral of Eq.~\eqref{eq:z_integral}, obtaining % 

\begin{align}
p(\mbh|\cbl,\xbl,\mg,\Theta)
&=|2\pi\Sigma_m|^{-\frac{1}{2}} \nonumber \\
&\times \exp \left[-\frac{1}{2}(\mbh -(\muu + \mg - \alpha \cdot \xbl + \beta \cdot \cbl))^T\Zm^{-1}(\mbh -(\muu + \mg - \alpha \cdot \xbl + \beta \cdot \cbl))\right] \label{eq:marginalized_z}
\end{align}
where from now on, $\muu = \muu(\zbh)$ and
\begin{align}
\Zm &=\Sigma_C+f\Sigma_z f^T \\
f &= \text{diag}(f_1, \dots, f_n) \\
f_i  &= 5\log_{10}(e) \left. \frac{D_{L}'(z_i)}{D_{L}(z_i)}\right|_{\hat{z_i}}\\
&= \frac{5\log_{10}(e)}{D_L(\hat{z}_i)} \left[\frac{D_L(\hat{z}_i)}{1+{z}_i} +\frac{c}{H_0}(1+\hat{z}_i)  \times \text{cosn} \{ \sqrt{|\Omk|}  \int_0^{\hat{z}} \dr z' 
 \[ (1+z')^3\Om + \Ode(z) + (1+z)^2\Omk \]^{-1/2} \} \right.\\
& \left. \times ( (1+z')^3\Om + \Ode(z) + (1+z)^2\Omk ]^{-1/2} ) \] 
\end{align}
Strictly speaking, one should integrate over redshift in the range $0 \leq z_i < \infty$, not  $-\infty < z_i < \infty$, which would result in the appearance of Gamma functions in the final result. However, as long as
$\frac{\sigma_{z_i}}{z_i}\ll1$ (as is the case here), this approximation is expected to be excellent.

% ********************************************************************************************************
\subsection{Integration over latent $\{ \cbl,\xbl,\mg\} $}
% ********************************************************************************************************

From Eq.~\eqref{eq:eff_like_1} and using the expression in Eq.~\eqref{eq:mobs}, we wish to integrate out the latent variables 
\begin{align}
Y&=\{ Y_1, \dots,Y_n\} \in \Real^{3n}, \\
Y_i&=\{ c_i, x_{1,i},M_i \}\in \Real^{3}, \\
\end{align}
We therefore recast expression \eqref{eq:mobs} as 
\begin{align}
p(\cbh,\xbh,\mbh|\cbl,\xbl,\mg,\Theta) &= |\Scov|^{-\hf} \exp \left( - \hf [(AY-X_0)^T\Scov^{-1}(AY-X_0)] \right) 
\end{align}
where we have defined the block-diagonal matrix 
\be
A = \text{diag}(T,T,\dots,T) \in \Real^{3n\times 3n}
\ee
with
\be 
T = 
\left[
\begin{array}{ccc}
1 &  0 & 0 \\
0 &  1 & 0 \\
\beta & -\alpha & 1
\end{array}
\right]
\left[
\begin{array}{c}
c_i \\
x_i \\
M_i
\end{array}
\right]
\ee
The prior terms appearing in Eq.~\eqref{eq:eff_like_2}, namely $p(\cbl | \cnot,\Rc) p(\xbl | \xnot,\Rs) p(\mg|M_0,\sigint)$, may be written as:
\begin{align}
p(\cbl | \cnot,\Rc)p(\xbl | \xnot,\Rs)p(\mg|M_0,\sigint) &=|2\pi \Sigp|^{-\hf}\exp \left( - \hf [(Y-Y_*)^T\Sigp^{-1}(Y-Y_*)] \right)  \notag \\
\end{align}
where%
\be
S^{-1}=\diag{\Rc^{-2}, \Rs^{-2},(\sigint)^{-2}} \in \Real^{3 \times 3}, \quad \Sigma_P^{-1} =\diag{S^{-1},S^{-1}, \dots,S^{-1}}    \in \Real^{3n \times 3n} ,
\ee
\begin{align}
Y_* &= \jj \cdot \bub \in \Real^{3n \times 1} , \\
\jj &= \left[ 
\begin{array}{ccc}
1 & 0 & 0 \\
0 & 1 & 0 \\
0 & 0 & 1 \\
\vdots & \vdots & \vdots \\
1 & 0 & 0 \\
0 & 1 & 0 \\
0 & 0 & 1 
\end{array}
\right] \in \Real^{3n \times 3}, \\
\bub &= \left[
\begin{array}{c}
c_* \\
x_* \\
M_0
\end{array}
\right] \in \Real^{3 \times 1} . 
\end{align}
Now the integral over $\dr Y =\dr \cbl \,\, \dr \xbl \,\, \dr \mg$ in Eq.~\eqref{eq:eff_like_2} can be performed, giving:
\begin{align}
\int  \dr Y  \,\,p(\cbh,\xbh,\mbh|\cbl,\xbl,\mg,\Theta)&p(\cbl | \cnot,\Rc)p(\xbl | \xnot,\Rs)p(\mg|M_0, \sigint) =  \notag \\ 
&= |2 \pi\Scov|^{-\hf} |2\pi \Sigp|^{-\hf} |2\pi \Sa|^{\hf}\exp \left( - \hf [X_0^T\Scov^{-1}X_0 -Y_0^T \Sa^{-1} Y_0 + Y_*^T \Sigp^{-1} Y_*] \right)    
\label{eq:likeintcxM}
\end{align}
where
\begin{align}
\Sa^{-1}&=A^T \Scov^{-1} A + \Sigp^{-1} \in \Real^{3n \times 3n} ,\\
\Sa^{-1}Y_0 &= A^T\Scov^{-1} X_0 + \Sigp^{-1}Y_* , \\
Y_0 &= \Sa (A^T\Scov^{-1} X_0 +  \Sigp^{-1}Y_*) \Sa (\Delta +  \Sigp^{-1}Y_*), \\
\Delta   &= A^T\Scov^{-1} X_0 \in \Real^{3n \times 1} .
\end{align}
Substituting  Eq.~\eqref{eq:likeintcxM} back into Eq.~\eqref{eq:eff_like_2} gives:
\begin{align}
p(\cbh,\xbh,\mbh|\Theta) &= \int\dr \Rc \, \, \dr \Rs \,\, \dr \cnot \,\, \dr \xnot \,\, |2 \pi\Scov|^{-\hf} |2\pi \Sigp|^{-\hf} |2\pi \Sa|^{\hf}\notag \\
& \times  \exp \left( - \hf [X_0^T\Scov^{-1}X_0 -Y_0^T \Sa^{-1} Y_0 + Y_*^T \Sigp^{-1} Y_*] \right)\notag \\
&\times p(\Rc)p(\Rs)p(\cnot)p(\xnot)p(M_0|\sigint). 
\label{eq:eff_nlike_2}
\end{align}
% ********************************************************************************************************
\subsection{Integration over population variables $\{ \cnot,\xnot, M_0 \}$}
% ********************************************************************************************************
The priors on the population variables $\bub = \{ \cnot,\xnot, M_0 \}$ in Eq.~\eqref{eq:eff_nlike_2} can be written as:
\begin{align}
p(\bub) = p(\cnot)p(\xnot)p(M_0|\sigint)&=|2 \pi \So|^{-\hf}\exp \left( -\hf  (\bub-\bub_m)^T \So^{-1} (\bub-\bub_m) \right)
\end{align}
where
\begin{align}
\So^{-1} &= \left[
\begin{array}{ccc}
1/\sigma^2_{c_*} & 0                & 0 \\
0                & 1/\sigma^2_{x_*} & 0 \\
0                & 0                &  1/\sigma^2_{M_0}
\end{array}
\right]
\end{align}
and
\begin{align}
\bub_m &= \left[
\begin{array}{c}
0 \\
0 \\
M_m
\end{array}
\right] \in \Real^{3 \times 1}
\end{align}
Thus Eq.~\eqref{eq:eff_nlike_2} can be written as:
\begin{align}
p(\cbh,\xbh,\mbh|\Theta)&= \int\dr \Rc \, \, \dr \Rs \,\, \dr \bub  \,\, |2 \pi\Scov|^{-\hf} |2\pi \Sigp|^{-\hf} |2\pi \Sa|^{\hf} |2\pi \So|^{-\hf} p(\Rc)p(\Rs) \notag \\
& \times  \exp \left( - \hf [X_0^T\Scov^{-1}X_0 -(\Sa (\Delta +  \Sigp^{-1}\jj \cdot \bub) )^T \Sa^{-1}(\Sa (\Delta +  \Sigp^{-1}\jj \cdot \bub) )  + \bub^T \jj^T \Sigp^{-1} \jj \bub +(\bub-\bub_m)^T \So^{-1}(\bub-\bub_m) ] \right)\notag \\
&= \int\dr \Rc \, \, \dr \Rs \,\, |2 \pi\Scov|^{-\hf} |2\pi \Sigp|^{-\hf} |2\pi \Sa|^{-\hf} |2\pi \So|^{-\hf} p(\Rc)p(\Rs) \notag \\
& \times  \exp \left( - \hf [X_0^T\Scov^{-1}X_0 -\Delta^T\Sa \Delta -k_0^TK^{-1}k_0  +\bub_m^T \So^{-1}\bub_m] \right) 
\int \dr \bub\,\, \exp \left( - \hf [(\bub-k_0)^TK^{-1}(\bub-k_0)] \right)\notag \\
\label{eq:eff_nlike_3}
\end{align}
where
\begin{align}
K^{-1}&=-\jj^T \Sigp^{-1}\Sa \Sigp^{-1} \jj + \jj^T \Sigp^{-1}\jj + \So^{-1}  \in \Real^{3 \times 3},\\
K^{-1}k_0 &= \jj^T \Sigp^{-1}\Sa \Delta +\So^{-1}\bub_m  \in \Real^{3 \times 1} ,\\
k_0 &= K(\jj^T \Sigp^{-1}\Sa \Delta +\So^{-1}\bub_m).
\end{align}
We can now carry out the Gaussian integral over $\bub$ in Eq.~\eqref{eq:eff_nlike_2}, obtaining our final expression for the effective likelihood, 
\begin{align}
p(\cbh,\xbh,\mbh|\Theta)&= \int \dr \log \Rc \,\,\dr \log \Rs \,\,  |2 \pi\Scov|^{-\hf} |2\pi \Sigp|^{-\hf} |2\pi \Sa|^{\hf} |2\pi \So|^{-\hf} |2\pi K|^{\hf} \notag \\
& \times  \exp \left( - \hf [X_0^T\Scov^{-1}X_0 -\Delta^T\Sa \Delta -k_0^TK^{-1}k_0 +\bub_m^T \So^{-1}\bub_m] \right), 
\label{eq:eff_like_final}
\end{align}
where we have chosen an improper Jeffreys' prior for the scale variables $\Rc, \Rs$: 
\be
p(\Rc) \propto \Rc^{-1} \Rightarrow p(\Rc) \dr \Rc \propto \dr \log \Rc,
\ee
and analogously for $\Rs$. These two remaining nuisance parameters cannot be integrated out analytically, so they need to be marginalized numerically. Hence, $\Rc, \Rs$ are added to our parameters of interest and are sampled over numerically, and then marginalized out from the joint posterior.

\end{document}